\input{aipcheck}

\documentclass[
    ,final            % use final for the camera ready runs
  ]
  {aipproc}

\layoutstyle{6x9}

\usepackage{color,amsmath,latexsym,amssymb,enumerate}
\usepackage{graphicx}
\usepackage{epstopdf}

\newcommand{\rhob}{\rho_b}
\newcommand{\rhobt}{\tilde{\rho}_b}
\newcommand{\vb}{{\bf v}_b}
\newcommand{\vbt}{\tilde{\bf v}_b}
\newcommand{\nablat}{\tilde{\nabla}}
\renewcommand{\(}{\left(}
\renewcommand{\)}{\right)}
\newcommand{\mn}{{\tt n}}
\newcommand{\xvec}{{\bf x}}
\newcommand{\Enu}{E_{\nu}}
\newcommand{\eh}{e_h}
\newcommand{\ec}{e_c}
\newcommand{\dt}{\Delta t}

\begin{document}

\title{Cosmological Radiation Hydrodynamics with {\em Enzo}}


\author{Michael L. Norman}{
  address={Physics Department, U.C.~San Diego, La Jolla, CA 92093}
  ,altaddress={Ctr. for Astrophysics and Space Sciences, U.C.~San
    Diego, La Jolla, CA 92093} % additional visiting address
}

\author{Daniel R. Reynolds}{
  address={Mathematics, Southern Methodist University, Dallas, TX 75275-0156}
}

\author{Geoffrey C. So}{
  address={Physics Department, U.C.~San Diego, La Jolla, CA 92093}
}

\begin{abstract}
  We describe an extension of the cosmological hydrodynamics code {\em Enzo} to include the self-consistent transport of ionizing radiation modeled in the flux-limited diffusion approximation. A novel feature of our algorithm is a coupled implicit solution of radiation transport, ionization kinetics, and gas photoheating, making the timestepping for this portion of the calculation resolution independent. The implicit system is coupled to the explicit cosmological hydrodynamics through operator splitting and solved with scalable multigrid methods. We summarize the numerical method, present a verification test on cosmological Str\"{o}mgren spheres, and then apply it to the problem of cosmological hydrogen reionization. 
\end{abstract}

\maketitle

%%%%%%%%%%%%%%%%%%%%%%%%%%%%%%%%%%%%%%%%%%%%
%% MAINMATTER
%%%%%%%%%%%%%%%%%%%%%%%%%%%%%%%%%%%%%%%%%%%%

\section{Introduction}

%<M.N.>

%[1 pg; motivation, relation to other methods]

It is a distinct privilege to help celebrate Dimitri Mihalas' 70th birthday by reporting on our latest work at this Festschrift in his honor. As a former faculty colleague of his at UIUC and scientific collaborator, I could tell many anecdotes. However of the many things I learned from Dimitri, the two that stick with me are: (1) solve the problem you need to solve to do the science, and (2) be rigorous about how you do it. These lessons motivate our present approach to self-consistent cosmological radiation hydrodynamics which is documented more fully in \cite{ReynoldsEtAl2009}.  

A current frontier in cosmological structure formation simulations is including the feedback of radiating sources such as galaxies and AGN on the intergalactic medium (IGM) in a self-consistent way. For example, the collective UV radiation from protogalaxies is believed to reionize the IGM at $z \approx 7-8$ \cite{2006ARA&A..44..415F}. This process can be thought of as the expansion and eventual overlap of R-type ionization fronts driven by high rates of star formation in the protogalaxies. R-type ionization fronts couple to the gas very weakly. Consequently a number of studies have simulated cosmological reionization by post-processing density fields taken from cosmological simulations with a standalone radiative transfer code ; e.g. \cite{2003MNRAS.344L...7C,2004MNRAS.350...47S,2006MNRAS.369.1625I}. However, closer to the sources and within the galaxies themselves where the gas density is higher, or when an intergalactic R-type front sweeps over a dense clump, ionization fronts may become D-type. When this happens coupling to gas motions is strong and a self-consistent approach to modeling is required in which hydrodynamics, radiative transfer, and the thermal/ionization state of the gas are evolved in a coupled fashion \cite{2006ApJS..162..281W}. 

\cite{2009arXiv0905.2920I} summarizes a variety of methods currently under development by the numerical cosmology community that do this. The radiative transfer methods employed include ray tracing, Monte Carlo, and moment methods. Here we present a method based on flux-limited diffusion. A novel feature of our algorithm is a coupled implicit solution of radiation transport, ionization kinetics, and gas photoheating, making the timestepping for this portion of the calculation resolution independent. This will be essential when adaptive mesh refinement (AMR) is employed. At present our algorithm is only implemented on uniform (non-adaptive) Cartesian grids. After describing our method, we verify its correctness on a cosmological Str\"{o}mgren sphere test problem. We then present a low-resolution ``first light" test of the coupled code to the problem of cosmological reionization. A more comprehensive description of our method and verification tests can be found in \cite{ReynoldsEtAl2009}.

\section{Physical Model}

We consider the coupled system of partial differential equations
\begin{align}
  \label{eq:cons_mass}
  \partial_t \rhob + \frac1a \vb \cdot \nabla
    \rhob &= -\frac1a \rhob \nabla\cdot\vb, \\
  \label{eq:cons_momentum}
  \partial_t \vb + \frac1a\(\vb\cdot\nabla\)\vb &=
    -\frac{\dot{a}}{a}\vb - \frac{1}{a\rhob}\nabla p - \frac1a
    \nabla\phi, \\
  \label{eq:cons_energy}
  \partial_t e + \frac1a\vb\cdot\nabla e &=
    - \frac{2\dot{a}}{a}e
    - \frac{1}{a\rhob}\nabla\cdot\left(p\vb\right) 
    - \frac1a\vb\cdot\nabla\phi + G - \Lambda \\
  \label{eq:chemical_ionization}
  \partial_t \mn_i + \frac{1}{a}\nabla\cdot\(\mn_i\vb\) &=
    \alpha_{i,j} \mn_e \mn_j - \mn_i \Gamma_{i}^{ph}, \qquad
    i=1,\ldots,N_s, \\
  \label{eq:cons_radiation}
  \partial_t E + \frac1a \nabla\cdot\(E \vb\) &= 
    \nabla\cdot\(D\nabla E\) - \frac{\dot{a}}{a}E + 4\pi\eta - c
    \kappa E, \\
  \label{eq:gravity}
  \nabla^2\phi &= \frac{4\pi g}{a}(\rhob + \rho_{dm} - \langle \rho \rangle).
\end{align}
These describe conservation of mass \eqref{eq:cons_mass},
conservation of momentum \eqref{eq:cons_momentum}, conservation of
energy \eqref{eq:cons_energy}, chemical rate equations
\eqref{eq:chemical_ionization}, flux-limited diffusion radiative
transfer \eqref{eq:cons_radiation} and Poisson's equation
\eqref{eq:gravity}, in a coordinate system that is comoving with the
expanding universe
\cite{BryanEtAl1995,HayesNorman2003,Paschos2005,ReynoldsEtAl2009}.   
The independent variables in these equations consist of the comoving
baryonic density $\rhob$, the proper peculiar baryonic velocity
$\vb$, the total gas energy per unit mass $e$, the comoving number
density for each chemical species $\mn_i,\; i=1,\ldots,N_s$, the
comoving radiation energy density $E$, and the modified gravitational
potential $\phi$.  

The cosmological flux-limited diffusion (FLD) equation
\eqref{eq:cons_radiation} deserves some comment.  In deriving this
equation from the general multi-frequency version \cite{Paschos2005}, 
\begin{equation}
\label{eq:cons_radiation_multifreq}
     \partial_{t} \Enu + \frac1a \nabla\cdot(\Enu\vb) =
     \nabla\cdot(D\,\nabla\Enu) + \nu \frac{\dot{a}}{a}\partial_{\nu}\Enu 
     + 4\pi\eta_{\nu} - c \kappa_{\nu} \Enu,
\end{equation}
we have assumed a prescribed radiation frequency spectrum,
$\chi_E(\nu)$, allowing the frequency-dependent radiation energy density
to be written in the form
$\Enu(\xvec,t,\nu)=\tilde{E}(\xvec,t)\,\chi_E(\nu)$.  With this
assumption, the single ``grey'' radiation energy density is given by
\begin{equation}
\label{eq:grey_definition}
  E(\xvec,t) = \int_{\nu_0}^{\infty} \Enu(\xvec,t,\nu) d\nu =
  \tilde{E}(\xvec,t) \int_{\nu_0}^{\infty} \chi_E(\nu)d\nu,
\end{equation}
and the equation \eqref{eq:cons_radiation} is then obtained through
integration of \eqref{eq:cons_radiation_multifreq} over frequencies
ranging from the ionization threshold of hydrogen ($h \nu_0=13.6$ eV)
to infinity.  In this paper we assume the radiation has a $T_B=10^5$
blackbody spectrum, i.e.~$\chi_E(\nu) = 8\pi h\left(\frac{\nu}{c}\right)^3
\big/\left(\exp\left(\frac{h\nu}{k_bT_B}\right)-1\right)$.  

The dependent variables in these equations are the proper pressure $p$,
the temperature $T$, and the comoving electron number density $\mn_e$,
given through the equations
\begin{align}
\label{eq:eos}
%  e &= \frac{p}{\rhob(\gamma-1)} + \frac12|\vb|^2, \\
  p &= \rhob(\gamma-1)\left(e - \frac12|\vb|^2\right), \\
\label{eq:temperature}
  T &= (\gamma-1)\frac{p\,\mu\, m_p}{\rhob\, k_b}, \\
\label{eq:electron_density}
  \mn_e &= \begin{cases} \mn_{HII}, &\text{(hydrogen only)} \\
                         \mn_{HII} + \tfrac14 \mn_{HeII} + \tfrac12
                         \mn_{HeIII}, &\text{(hydrogen + helium)}.
                       \end{cases}
\end{align}
Here $\gamma$ is the ratio of specific heats, which we take to be
$5/3$; $m_p$ corresponds to the mass of a proton and $k_b$ is
Boltzmann's constant.  The local molecular weight $\mu$ depends on the
density and chemical ionization state.

In addition, the system \eqref{eq:cons_mass}-\eqref{eq:gravity}
contains a number of coupling terms and coefficients.  The coefficient
$a(t)\equiv(1+z)^{-1}$ denotes the cosmological expansion parameter
for a smooth homogeneous background, where the redshift $z$ 
is a function of time only; all spatial derivatives are therefore
taken with respect to the comoving position $\xvec\equiv{\bf r}/a(t)$.
The term $\dot{a}=\frac{\mathrm{d} a(t)}{\mathrm{d}t}$.
$a(t)$ is obtained by integrating the Friedmann equation for the 
set of assumed cosmological parameters.
The gas heating and cooling rates $G$ and $\Lambda$ are functions of
the temperature, radiation energy density and chemical ionization
state.  The temperature-dependent chemical reaction rates
$\alpha_{i,j}$ define the interactions between chemical species, and
the photoionization rate $\Gamma_i^{ph}$ depends on the radiation
energy density.  Formulas for all of these terms may be found in the 
references
\cite{Black1981,Osterbrock1989,Cen1992,AbelEtAl1997,HuiGnedin1997,RazoumovEtAl2002}. 

In the radiation equation \eqref{eq:cons_radiation}, $c$ is the speed
of light, the total opacity $\kappa$ is a function of the chemical
ionization state, and the emissivity $\eta$ is provided as either a
radiation source term, or may depend on the density, velocity, gas
energy, and chemical ionization state.  The formulae for these
dependencies may be found in the references
\cite{Osterbrock1989,1992ApJ...399L.113C}.  Of special importance in
this equation is the coefficient function $D$, which in a flux-limited
diffusion approximation attempts to allow the equation to span
behaviors ranging from nearly isotropic to free-streaming radiation.
To this end, we choose the coefficient to be of the form
\begin{equation}
\label{eq:limiter}
   D(E) = \text{diag}\(D_1(E), D_2(E), D_3(E)\),
   \quad\text{where}\quad 
   D_i(E) = \frac{c(2\kappa_T+R_i)}{6\kappa_T^2+3\kappa_T R_i+R_i^2},
\end{equation}
with $R_i = |\partial_i E|/E,\, i=1,2,3$.  Here
$\kappa_T=\kappa+\kappa_S$ is the total extinction coefficient, 
where $\kappa$ is the opacity and $\kappa_S$ results from scattering
\cite{HayesEtAl2006}.  The function \eqref{eq:limiter} has been
reformulated from its original version \cite{HayesNorman2003} to
provide increased stability for scattering-free simulations involving
extremely small opacities (i.e.~$\kappa_T=\kappa\ll 1$), as
is typical in cosmology applications.

In the Poisson equation for the gravitational potential
\eqref{eq:gravity}, the baryonic gas is coupled to collisionless dark
matter $\rho_{dm}$ and the cosmic mean density $\langle \rho
\rangle$ solely through their self--consistent gravitational field.
Here, $g$ provides the gravitational constant, and the dark matter
density is evolved using the Particle-Mesh method described in
\cite{HockneyEastwood1988,NormanBryan1999,OSheaEtAl2004}.

\section{Algorithm}

Instead of working with the equations directly in CGS units, we first
normalize the system to render the values tractable for floating point
computation.  To this end, we define the scaled units
\begin{equation}
  \label{eq:units}
  \tilde{x} = x / u_x, \qquad 
  \tilde{g} = g / u_g, \qquad 
  \tilde{t} = t / u_t, 
\end{equation}
where the constants $u_x$, $u_t$ and $u_g$ correspond to
the typical magnitudes of length, time and mass at the start of the
simulation.  We further define the density unit factor $u_d =
u_g/(u_x)^3$ and velocity scaling factor $u_v = u_x/u_t$.  We note
that due to our use of {\em comoving} length, these constants are all
redshift-independent.  The {\em proper} length values at any point in
the simulation are therefore given by
\begin{equation}
  \label{eq:proper_length}
  x_{\text{proper}} = x\, a(t) = \tilde{x}\, u_x\, a(t).
\end{equation}
With these unit scalings, we define the normalized variables
\begin{align}
  \label{eq:scalings}
  \rhobt = \rho / u_d, &\qquad \vbt = \vb / u_v, \qquad
  \tilde{e} = e / u_v^2, \\
  \notag
  \tilde{E} = E / \(u_d u_v^2\), &\qquad \tilde{\mn}_i = \mn_i / u_d,
  \qquad \tilde{\phi} = \phi /u_v^2.
\end{align}
The proper densities may be obtained from the comoving densities
through the formulae 
\begin{align}
  \notag
  E_{\text{proper}} &= E / a^3(t) = \tilde{E} \frac{u_E}{a^3(t)}, \\
  \label{eq:proper_vars}
  \mn_{i,{\text{proper}}} &= \mn_i / a^3(t) = \tilde{\mn}_i \frac{u_n}{a^3(t)}, \\
  \notag
  \rho_{b,\text{proper}} &= \rhob / a^3(t) = \rhobt \frac{u_n}{a^3(t)}.
\end{align}
With these rescaled variables, we rewrite our equations
\eqref{eq:cons_mass}-\eqref{eq:gravity} as the normalized system
\begin{align}
  \label{eq:cons_mass_rescaled}
  \partial_{\tilde{t}} \rhobt + \frac1a \vbt \cdot \nablat
    \rhobt &= -\frac1a \rhobt \nablat\cdot\vbt, \\
  \label{eq:cons_momentum_rescaled}
  \partial_{\tilde{t}} \vbt + \frac1a\(\vbt\cdot\nablat\)\vbt &=
    -\frac{\dot{a}}{a}\vbt - \frac{1}{a\rhobt}\nablat \tilde{p} - \frac1a
    \nablat\tilde{\phi}, \\
  \label{eq:cons_energy_rescaled}
  \partial_{\tilde{t}} \tilde{e} + \frac1a\vbt\cdot\nablat \tilde{e} &=
    - \frac{2\dot{a}}{a}\tilde{e}
    - \frac{1}{a\rhobt}\nablat\cdot\left(\tilde{p}\vbt\right) 
    - \frac1a\vbt\cdot\nablat\tilde{\phi} + \tilde{G} - \tilde{\Lambda} \\
  \label{eq:chemical_ionization_rescaled}
  \partial_{\tilde{t}} \tilde{\mn}_i +
    \frac{1}{a}\nablat\cdot\(\tilde{\mn}_i\vbt\) &= 
    \tilde{\alpha}_{i,j} \mn_e \tilde{\mn}_j - \tilde{\mn}_i
    \tilde{\Gamma}_{i}^{ph}, \qquad 
    i=1,\ldots,N_s, \\
  \label{eq:cons_radiation_rescaled}
  \partial_{\tilde{t}} \tilde{E} + \frac1a \nablat\cdot\(\tilde{E} \vbt\) &= 
    \nablat\cdot\(D\nablat \tilde{E}\) - \frac{\dot{a}}{a}\tilde{E} +
    4\pi\tilde{\eta} - c \tilde{\kappa} \tilde{E}, \\
  \label{eq:gravity_rescaled}
  \nablat^2\tilde{\phi} &= \frac{4\pi \tilde{g}}{a}(\rhobt + \tilde{\rho}_{dm} -
  \langle \tilde{\rho} \rangle). 
\end{align}
Here, $\dot{a}$ now refers to the derivative 
$\frac{\mathrm{d} a}{\mathrm{d}\tilde{t}}$.  For clarity of notation,
all subsequent variables are shown without the $\sim$ superscript,
although all solver algorithms operate on the normalized variables.

\subsection{Operator-Split Hydrodynamics with Radiative Feedback}
\label{sec:operator_split}

We solve the coupled system
\eqref{eq:cons_mass_rescaled}-\eqref{eq:gravity_rescaled} 
using an operator-split framework, wherein we solve sub-components of
the system one at a time, feeding the results of each sub-system into
the remaining parts.  In this approach, a time step is taken with the
steps: 
\begin{itemize}
\item[(i)] Project the dark matter particles onto the finite-volume
  mesh to generate the dark matter density field $\rho_{dm}$.
\item[(ii)] Solve for the gravitational potential $\phi$ using
  equation \eqref{eq:gravity_rescaled}. 
\item[(iii)] Advect the dark matter particles with the Particle-Mesh
  method \cite{HockneyEastwood1988,NormanBryan1999,OSheaEtAl2004}.  
\item[(iv)] Evolve the hydrodynamics equations
  \eqref{eq:cons_mass_rescaled}-\eqref{eq:cons_energy_rescaled} with a
  high-order, explicit-time upwind method. In this step, the velocity
  field $\vb$ advects both the chemical number densities $\mn_i$ and
  radiation energy density $E$.
\item[(v)] Using a high-order implicit method, solve a coupled
  reaction-diffusion system
  \eqref{eq:cons_energy_rescaled}-\eqref{eq:cons_radiation_rescaled}
  to obtain the updated number densities $\mn_i$, radiation $E$ and
  gas energy $e$. 
\end{itemize}
The equation \eqref{eq:cons_energy_rescaled} is involved in both steps (iv) and
(v) above.  To do this, we split the gas energy into two parts,
$e=\eh+\ec$, where $\eh$ results from the hydrodynamic evolution of
the system (step (iv)), and $\ec$ is the gas energy {\em correction}
that results from couplings with radiation and chemistry (step (v)).
With this splitting, the hydrodynamic solver used in step (iv) of the
algorithm solves the system of equations 
\begin{align}
  \label{eq:enzo-cons_mass}
  \partial_t \rhob + \frac1a \vb \cdot \nabla
    \rhob &= -\frac1a \rhob \nabla\cdot\vb, \\
  \label{eq:enzo-cons_momentum}
  \partial_t \vb + \frac1a\(\vb\cdot\nabla\)\vb &=
    -\frac{\dot{a}}{a}\vb - \frac{1}{a\rhob}\nabla p - \frac1a
    \nabla\phi, \\
  \label{eq:enzo-cons_energy}
  \partial_t \eh + \frac1a\vb\cdot\nabla \eh &=
    - \frac{2\dot{a}}{a}\eh
    - \frac{1}{a\rhob}\nabla\cdot\left(p\vb\right) 
    - \frac1a\vb\cdot\nabla\phi \\
  \label{eq:enzo-chemical_ionization}
  \partial_t \mn_i + \frac{1}{a}\nabla\cdot\(\mn_i\vb\) &= 0, \\
  \label{eq:enzo-cons_radiation}
  \partial_t E + \frac1a \nabla\cdot\(E \vb\) &= 0,
\end{align}
using the Piecewise Parabolic Method (PPM) \cite{ColellaWoodward1984}, 
on a regular finite-volume spatial grid.  This solve evolves
$(\rhob^n,\vb^n,e^n,\mn_i^n,E^n)$ to the time-updated variables
$(\rhob^{n+1},\vb^{n+1},\eh^{n+1})$ and the advected variables
$(\mn_i^*,E^*)$, and is implemented in the community astrophysics code
Enzo \cite{OSheaEtAl2004,NormanEtAl2007,enzo-site}. 

Step (v) then solves the coupled system,
\begin{align}
  \label{eq:RT-cons_energy}
  \partial_t \ec &= -\frac{2\dot{a}}{a}\ec
    + G - \Lambda, \\
  \label{eq:RT-chemical_ionization}
  \partial_t \mn_i &= \alpha_{i,j} \mn_e \mn_j - \mn_i \Gamma_{i}^{ph}, \\
  \label{eq:RT-cons_radiation}
  \partial_t E &= \nabla\cdot\(D\nabla E\) - m\frac{\dot{a}}{a}E +
    4\pi\eta - c \kappa E, 
\end{align}
using a fully implicit nonlinear solution approach to evolve the
advected variables $(e_c^n,\mn_i^*,E^*)$ to the time-evolved 
quantities $(e_c^{n+1},\mn_i^{n+1},E^{n+1})$.  To this end, we define
the vector of unknowns $U=(e_c,\mn_i,E)^T$, and write the nonlinear
residual function $f(U) = (f_e,f_{\mn_i},f_E)^T$, where
\begin{align}
\label{eq:gas_residual}
  f_e(U) \ \equiv \ & e_c - S_e(U), \\
\label{eq:chemistry_residual}
  f_{\mn_i}(U) \ \equiv \ & \mn_i - S_{\mn_i}(U), \quad i=1,\ldots,N_s, \\
\label{eq:radiation_residual}
  f_E(U) \ \equiv \ & E - E^n - \dt\,\theta\left(
    \nabla\cdot\(D\nabla E\) - \frac{\dot{a}}{a} E + 4\pi\eta - c k
    E\right) \\
  \notag
    & - \dt(1-\theta)\left(\nabla\cdot\(D^n\nabla E^n\) - \frac{\dot{a}}{a}
    E^n + 4\pi\eta^n - c k^n E^n\right),
\end{align}
Here the functions $S_e(U)$ and $S_{\mn_i}(U)$ provide the
analytical solutions to an $O(\Delta t^2)$-accurate approximation
of the spatially-local ODE system
\eqref{eq:RT-cons_energy}-\eqref{eq:RT-chemical_ionization} for a
given value of $E$ \cite{Reynolds2009}.  The residual equation
\eqref{eq:radiation_residual} defines a standard two-level $\theta$
method for time integration of the equation
\eqref{eq:RT-cons_radiation}.  Therefore this overall nonlinear 
residual defines an up-to-second order implicit time discretization of
the coupled system
\eqref{eq:RT-cons_energy}-\eqref{eq:RT-cons_radiation}; where the 
time-evolved state $U^{n+1}$ is found through solution of the problem
$f(U)=0$.

To solve this nonlinear problem, we use a {\em globalized Inexact
Newton's Method} \cite{Kelley1995,KnollKeyes2004}, that iteratively
proceeds toward the solution $U^{n+1}$ through approximately solving a
sequence of linearized problems $J(U_k) S_k = -f(U_k)$, where $J(U_k)$
is the {\em Jacobian} of the nonlinear function $f$, evaluated at the
current Newton iterate $U_k$.  A full description of our solution
algorithm is provided in \cite{ReynoldsEtAl2009}.  To summarize this
process, we solve these linear Newton systems through a Schur 
complement formulation, that reduces the coupled linear system 
to a sequence of simpler sub-systems, culminating in an update to a
modified radiation equation.  This Schur complement subsystem is  
solved using a multigrid-preconditioned conjugate gradient method from
the HYPRE library \cite{FalgoutYang2002,hypre-site,BakerFalgoutYang2006}.

We measure convergence of the Newton iteration with the RMS norm
\begin{equation}
\label{eq:rmsnorm}
  \|w\| = \left(\frac{\|w\|_2^2}{N(N_s+2)}\right)^{1/2},
\end{equation}
where $N(N_s+2)$ is the number of unknowns in $w$, since this norm
does not grow artificially larger with mesh refinement.

\section{Verification tests}

The model equations and solution algorithm in this paper have been
rigorously tested on a myriad of problems, ranging from pure radiation
transport, to interacting radiation and hydrodynamics, to dynamic
radiation-hydrodynamics with chemical ionization
\cite{ReynoldsEtAl2009}.  In lieu of reiterating those tests
here, we demonstrate the approach on a single problem before moving on
to our target application.

We consider a verification problem that performs isothermal ionization
of a static (i.e., no fluid motions other than Hubble expansion) 
neutral hydrogen region, within a cosmologically expanding
universe.  The problem is originally due to Shapiro \& Giroux
\cite{ShapiroGiroux1987}, and exercises the radiation transfer,
cosmology and chemical ionization components of the coupled solver.  
The physics of interest in this example is the expansion of an ionized
hydrogen region in a uniform gas around a single monochromatic source,
emitting $\dot{N}_{\gamma} = 5\times 10^{48}$ photons per second at
the ionization frequency of hydrogen ($h\nu = 13.6$ eV).  Given the
initially-neutral hydrogen region and strength of the ionizing source,
the ionization region expands rapidly at first, with the I-front
approaching the equilibrium position where ionizations and
recombinations balance, referred to as the Str{\"o}mgren radius.
However, due to cosmological expansion, this equilibrium radius
begins to increase much faster than the I-front can propagate.  The
analytical formula for the location of the Str{\"o}mgren radius as a
function of time is
\begin{equation}
\label{eq:SG-stromgren_radius}
   r_S(t) = \left[\frac{3\dot{N}_{\gamma}}{4\pi \alpha_B
      \mn_H(t)^2}\right]^{1/3},
\end{equation}
where the proper hydrogen number density $\mn_H$ decreases due to
cosmological expansion by a factor of $a^{-3}(t)$.  Here
$\alpha_B\approx 2.59\times 10^{-13}$ cm$^3$/s is the case-B hydrogen
recombination coefficient.  If we define $\lambda = \alpha_B\mn_{H,0}
/ H_0 / (1+z_0)$, where $\mn_{H,0}$ is the hydrogen number density at
the initial redshift $z_0$, we may calculate the analytical I-front
radius at any point in time as 
\begin{align}
  \label{eq:SG-solution}
  r_I(t) &= r_{S,0} \left(\lambda\,e^{-\tau(t)}\int_{1}^{a(t)}
    e^{\tau(\tilde{a})}\left[1-2q_0+2q_0(1+z_0)/\tilde{a}\right]^{-1/2}
    \mathrm{d}\tilde{a}\right)^{1/3}, \\
  \notag \text{where}\qquad&\\
  \label{eq:SG-tau}
  \tau(a) &= \lambda\,\left[6\,q_0^2\,(1+z_0)^2\right]^{-1}\left[F(a)-F(1)\right], \\
  \label{eq:SG-F}
  F(a) &= \left[2 - 4q_0 - 2q_0(1+z_0)/a\right]
  \left[1-2q_0 + 2q_0(1+z_0)/a\right]^{1/2}, 
\end{align}
and where $q_0$ is the cosmological deceleration parameter. 

We perform four of the tests provided in the original paper
\cite{ShapiroGiroux1987}:  $q_0 = \{0.5, 0.05\}$, and $z_0 =
\{4,10\}$.  These correspond to the cosmological parameters
found in Table \ref{table:cosmology}.
\begin{table}
\begin{tabular}{lrccccc}
\hline
    \tablehead{1}{c}{b}{$q_0$}
  & \tablehead{1}{c}{b}{$z_0$}
  & \tablehead{1}{c}{b}{$L_0$}
  & \tablehead{1}{c}{b}{$H_0$}
  & \tablehead{1}{c}{b}{$\Omega_m$}
  & \tablehead{1}{c}{b}{$\Omega_{\Lambda}$}
  & \tablehead{1}{c}{b}{$\Omega_b$}   \\
\hline
   0.5 &  4 & 80 kpc & 0.5 & 1.0 & 0 & 0.2 \\
  0.05 &  4 & 60 kpc & 1.0 & 0.1 & 0 & 0.1 \\
   0.5 & 10 & 36 kpc & 0.5 & 1.0 & 0 & 0.2 \\
  0.05 & 10 & 27 kpc & 1.0 & 0.1 & 0 & 0.1 \\
\hline
\end{tabular}
\caption{Cosmological parameters for the verification tests.  See text
  for descriptions.}
\label{table:cosmology}
\end{table}
Here, $L_0$ is the initial box size and $H_0$ is the Hubble constant.
The values $\Omega_m$, $\Omega_{\Lambda}$ and $\Omega_b$ are the
contributions to the gas energy density at $z=0$ due to
non-relativistic matter, the cosmological constant, and baryonic
matter, respectively.  These two deceleration parameters result
in slightly different functions for the expansion coefficient $a$.
For $q_0=0.05$ we compute $a(t)$ using equations (13-3) and (13-10)
from \cite{Peebles1993}, while for $q_0=0.5$ we compute $a(t) =
(1+z(t))^{-1}$.  We begin all problems with an initial radiation
energy density of $E = 10^{-35}$ erg cm$^{-3}$ and an initial
ionization fraction $n_{HII}/\mn_{H,0}=0$.  
The initial density is chosen as
$\rho_{b,0}=1.175\times10^{-28}$ g cm$^{-3}$ for $q_0=0.5$, or
$\rho_{b,0}=2.35\times10^{-28}$ g cm$^{-3}$ for $q_0=0.05$.
All simulations are run from the initial redshift $z_0$ to $z=0$.
The ionization source is located in the lower corner of the box.  We
use reflecting boundary conditions at the lower 
boundaries and outflow conditions at the upper boundaries in each
direction.  The implicit solver used a convergence norm of $p=2$, time
step parameter of $\theta=0.51$, a desired temporal solution accuracy
$\tau_{\text{tol}}=0.001$ and inexactness parameter
$\delta_k=10^{-13}\|f(U_k)\|$ (see \cite{ReynoldsEtAl2009} for further
explanation of these parameters).   

\begin{figure}
 \centerline{
   \includegraphics[width=0.5\linewidth]{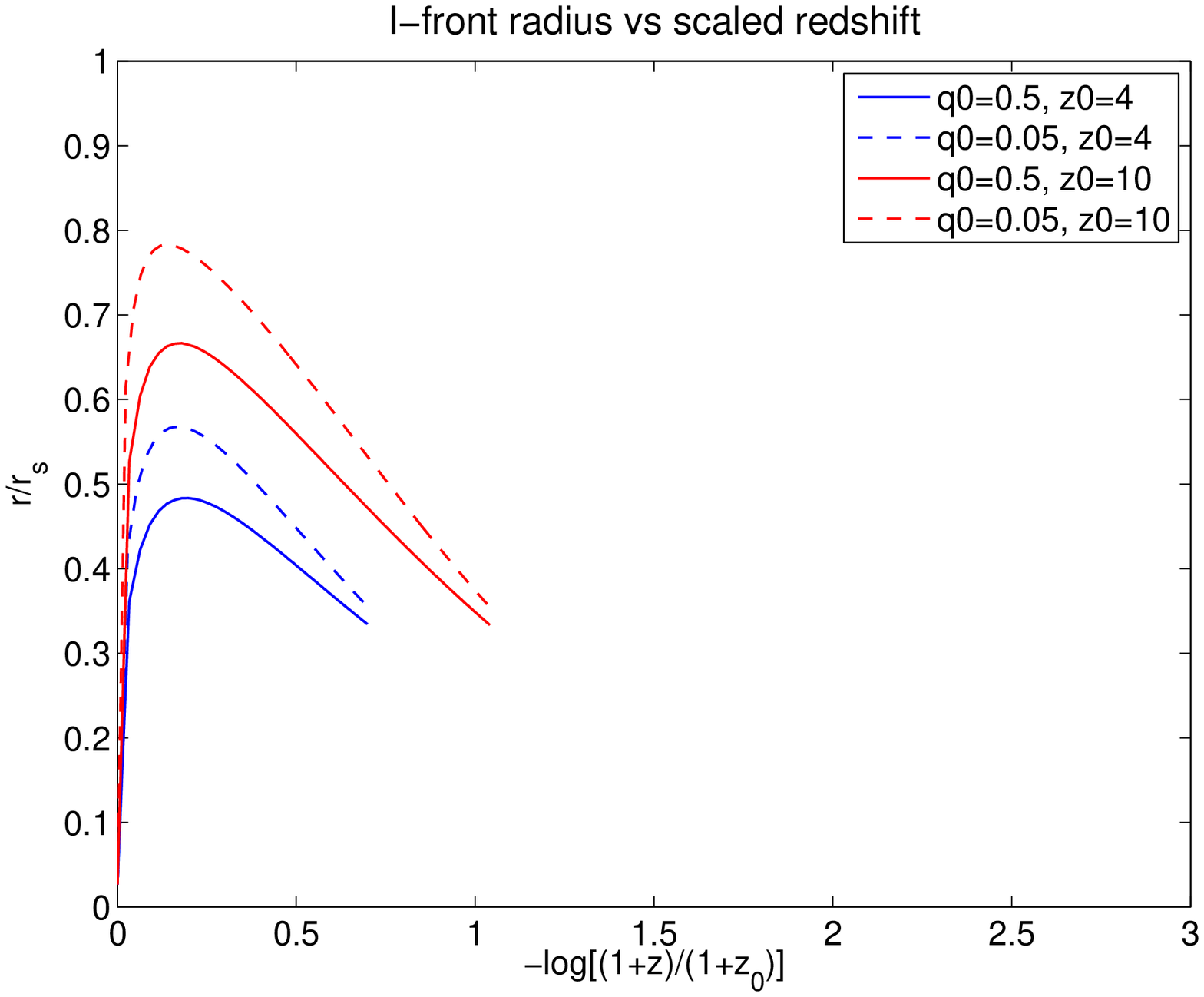}   
   \hfill
   \includegraphics[width=0.5\linewidth]{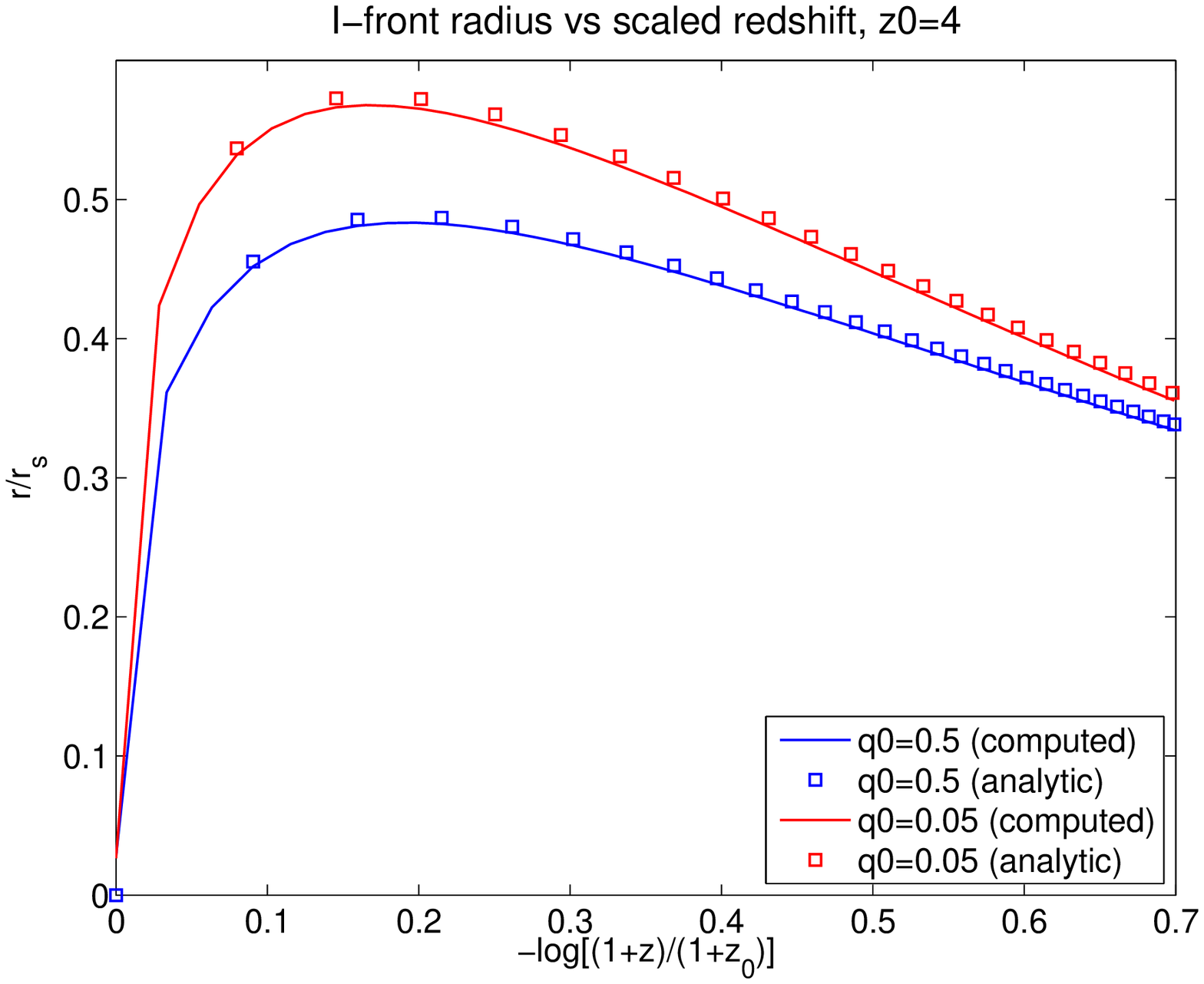}}
 \caption{Left: I-front radius vs.~scaled redshift for the four tests.
   Right: I-front radius vs scaled redshift for the $z_0=4$ tests;
   analytical solution values are shown with open squares.}
 \label{fig:ShapiroGiroux_evolution}
\end{figure}

\begin{figure}
 \centerline{
   \includegraphics[width=0.5\linewidth]{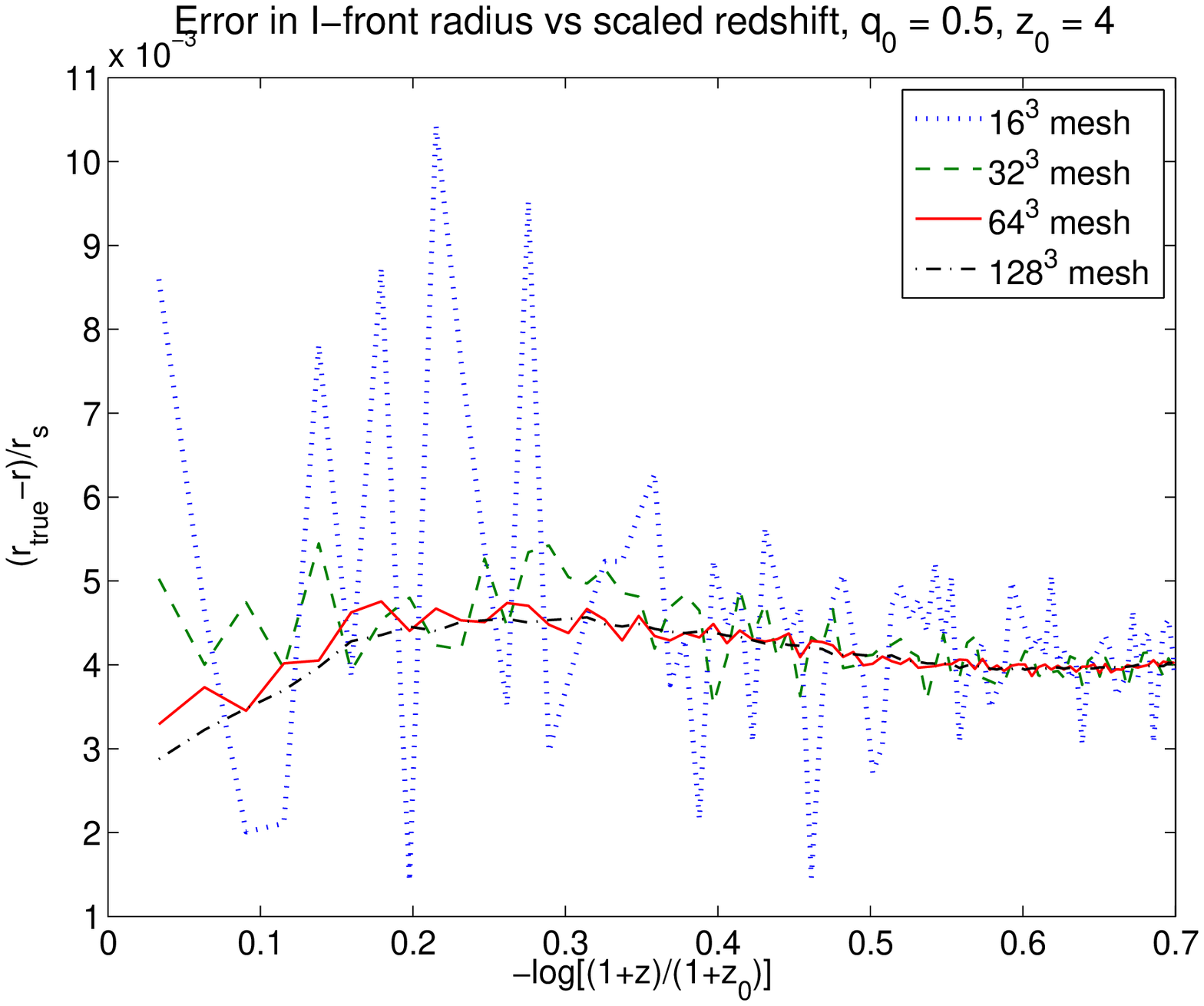}
   \hfill
   \includegraphics[width=0.5\linewidth]{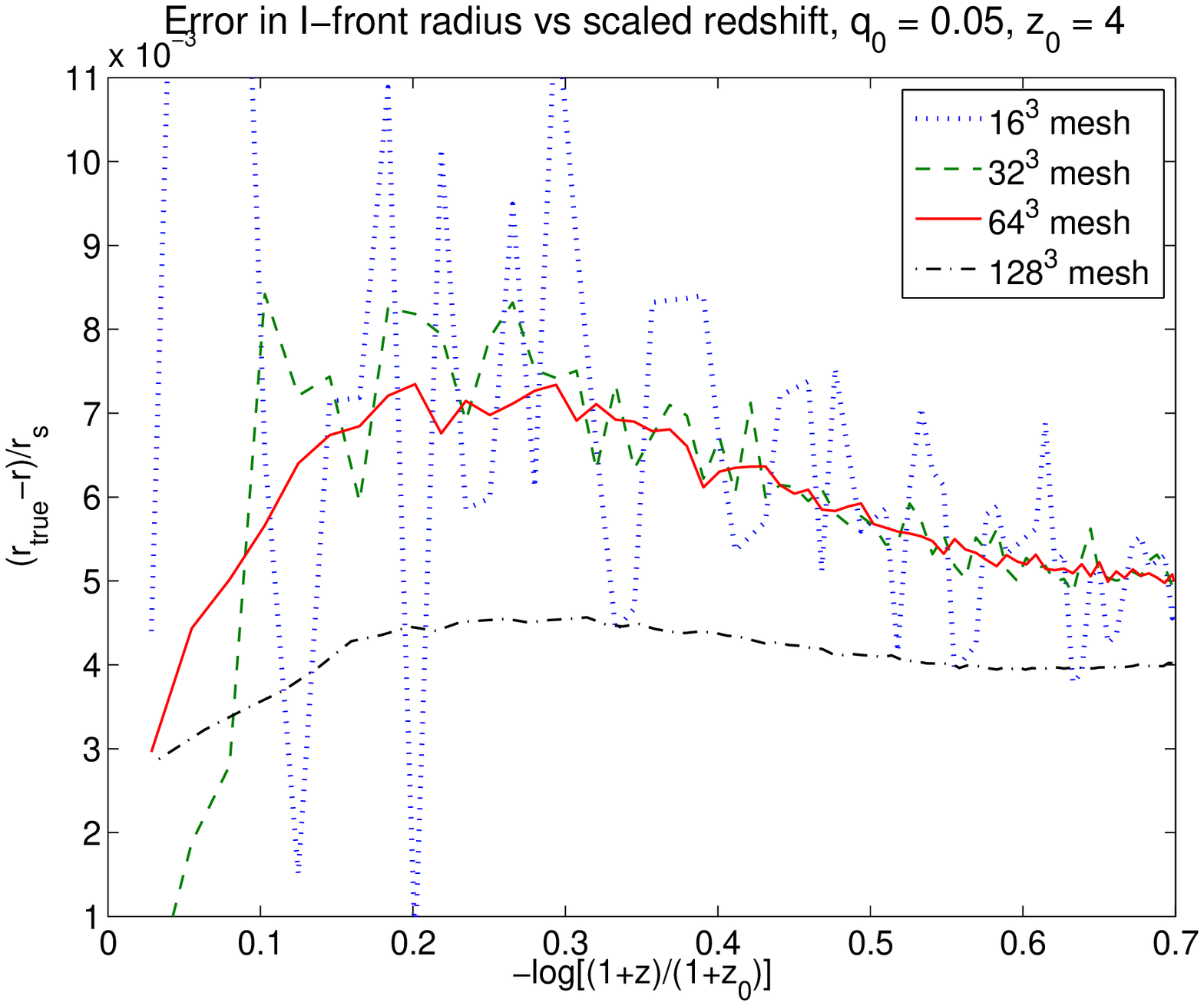}}
 \caption{Convergence of I-front radius vs. scaled redshift for the
   cases $q_0=\{0.5,0.05\}$ and $z_0=4$ as the mesh is refined: spatial
   meshes shown are $16^3$ (blue dotted), $32^3$ (green dashed),
   $64^3$ (red solid), and $128^3$ (black dot-dashed).} 
 \label{fig:ShapiroGiroux_convergence}
\end{figure}

In Figure \ref{fig:ShapiroGiroux_evolution} we plot the
scaled, spherically-averaged I-front position with respect to scaled
redshift for each of the four tests (with axes identical to
\cite{ShapiroGiroux1987}, Figure 1a), as well as the zoomed-in version
for the $z_0=4$ tests along with their analytical solutions; all of
these tests used a $128^3$ spatial mesh.
In Figure \ref{fig:ShapiroGiroux_convergence} we plot the error in the
computed I-front radius as we varied the spatial mesh size for the two
cases $(q_0,z_q)=(0.5,4)$ and $(0.05,4)$.  The accuracy in the
computed radius improves with mesh refinement.

\section{Application to cosmological reionization}

To illustrate the operation of the combined cosmological radiation hydrodynamics
plus ionization kinetics code, we simulate hydrogen reionization due to stellar
sources in a small cosmological volume. 
This is a low resolution functionality test only to show 
that the two halves of the code are coupled correctly;
scientific predictions will require considerably higher resolution and larger
boxes \cite{2008ApJ...689L..81T}. 

We simulate a $\Lambda$CDM cosmological model with the
following parameters: $\Omega_\Lambda = 0.7$, $\Omega_b = 0.04$,
$\Omega_{CDM} = 0.26$, $h=0.7$, $\sigma_8 = 0.9$, where these are, respectively,
the fraction of the closure density in vacuum energy, baryons, and cold dark
matter, the Hubble constant in units of 100 km/s/Mpc, and the variance of matter fluctuations in 8 Mpc$^{-1}$ spheres, all measured at the present epoch (z=0).
A Gaussian random field is initialized at z=99 using 
an initial power spectrum following \cite{1999ApJ...511....5E}.
The simulation was run in an 8 Mpc comoving box using a $64^3$ uniform grid and 
dark matter particles with periodic boundary conditions.  

The emissivity $\eta$ was computed
using a modified version of the star formation/feedback recipe of
\cite{1992ApJ...399L.113C},
in which the conditions for star formation within a computational cell
require that the local
$\vb$ have negative divergence (i.e~baryons are contracting), the radiative
cooling time is smaller than the dynamical time, and that $\rhob$ is
greater than some threshold (without checking for the Jeans mass).
If these criteria are met a star particle is created which represents an ensemble of
stars and
becomes a source of emissivity for the radiation solver. 
Many particles are created over time and there may be multiple star particles
in a cell. The emissivity of a cell $\eta$ is computed as follows:
\begin{equation}
  \label{eq:Emissivity}
  \eta = \frac{1}{4\pi}\sum\limits_i\, \epsilon_{UV} \int_t^{t+\Delta t} \dot{m}_{SF}(t)\, c^2\, \mathrm{d}t
\end{equation}
where the sum is over all the star particles in the cell, $\dot{m}_{SF}(t)$ is the star formation rate, which is an assumed analytic function of time for each particle, $c$ is the speed of light, $\Delta t$ is the timestep, and $\epsilon_{UV}$ an efficiency 
factor that
depends on a number of hidden parameters including the initial mass
function of the star cluster, the stellar spectral energy distribution, and the ionizing photon escape fraction. For simplicity, we used the upper value from
\cite{RazoumovEtAl2002} in its place.  

Snapshots from the simulation are created by the analysis tool {\em yt} \cite{SciPyProceedings_46} 
and shown in Table \ref{table:outputs}. Here projections through
the three dimensional volume are shown for four redshifts
$z$ = 4.35, 2.55, 1.99 and 0 and three physical quantities:
baryonic density ($\log_{10}(\rhob/\rho_{avg})$),
ionization fraction ($\log_{10}(\mn_{HII}/\mn_H)$) and temperature
($\log_{10}T$) in Kelvin.   

The first star particle was created at $z = 5.58$.
At this point, the initially homogeneous Intergalactic Medium (IGM)
had formed filamentry structures as a result of dark matter clumps.
By the first snapshot at $z = 4.35$, the effects of the star can
clearly be seen in the higher ionization fraction in the lower right
corner region around the star. This same region is evident by a
brighter peak in the density and temperature. 
\begin{table}
  \centering
    \begin{tabular}{lcr}
    \includegraphics[width=0.3\textwidth]{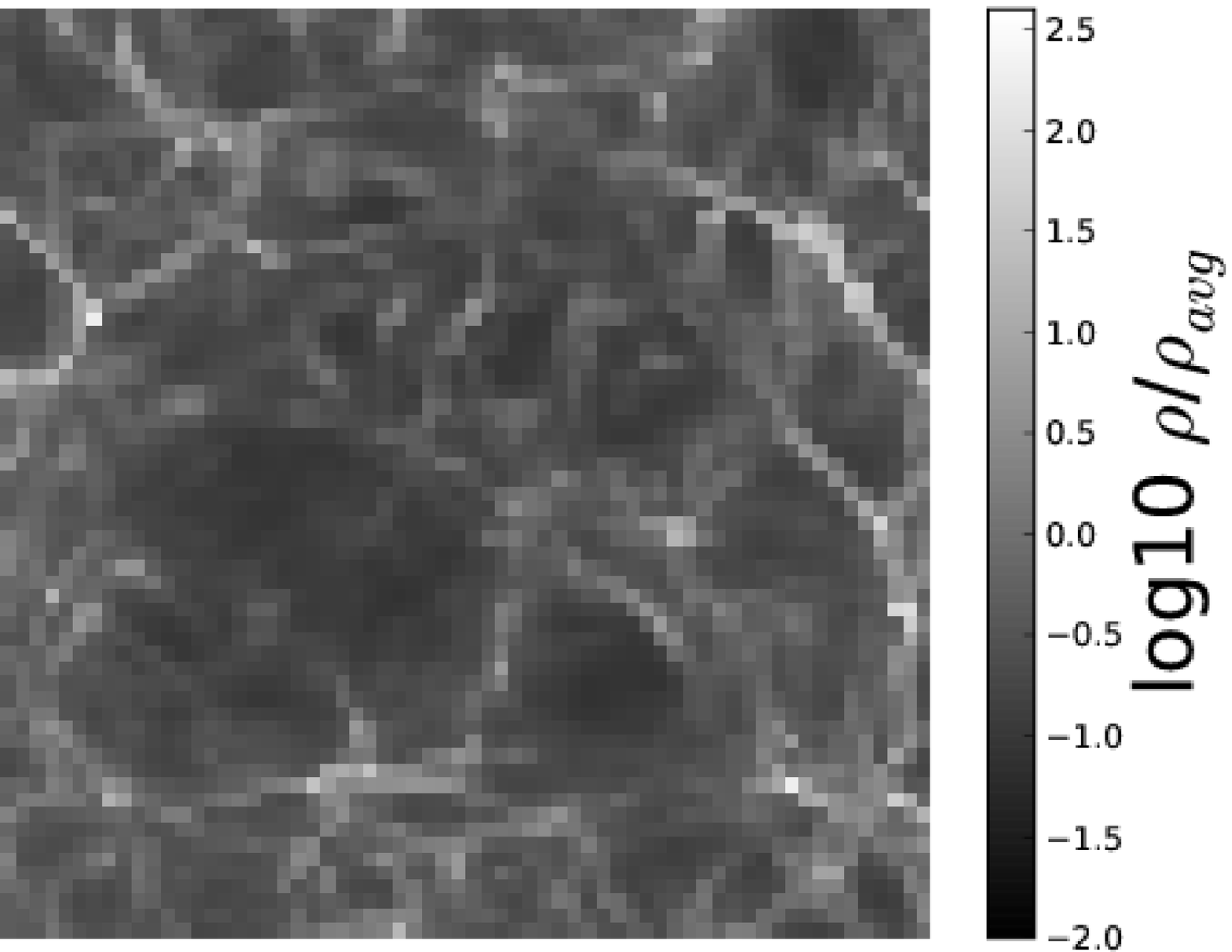} &
    \includegraphics[width=0.3\textwidth]{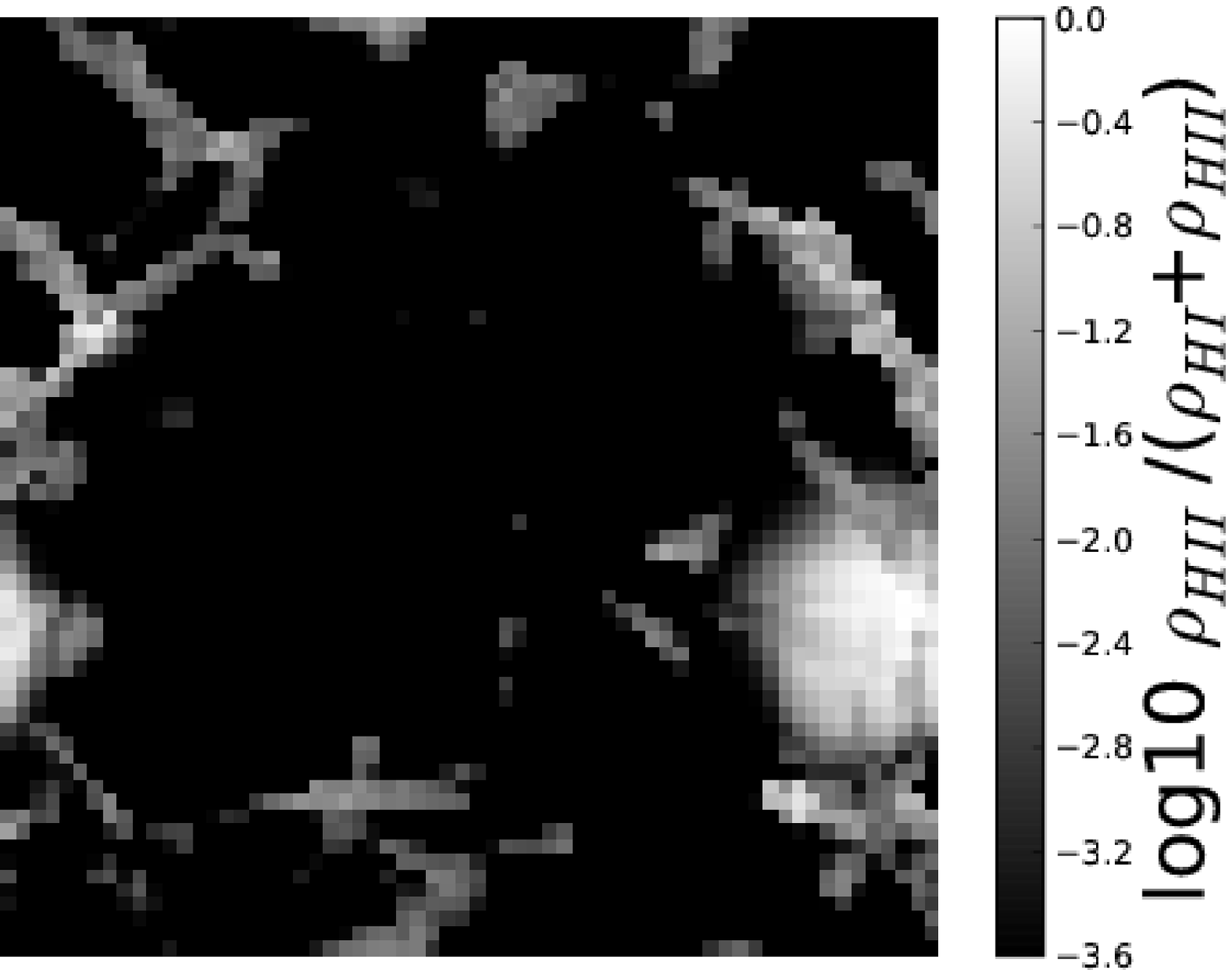} &
    \includegraphics[width=0.3\textwidth]{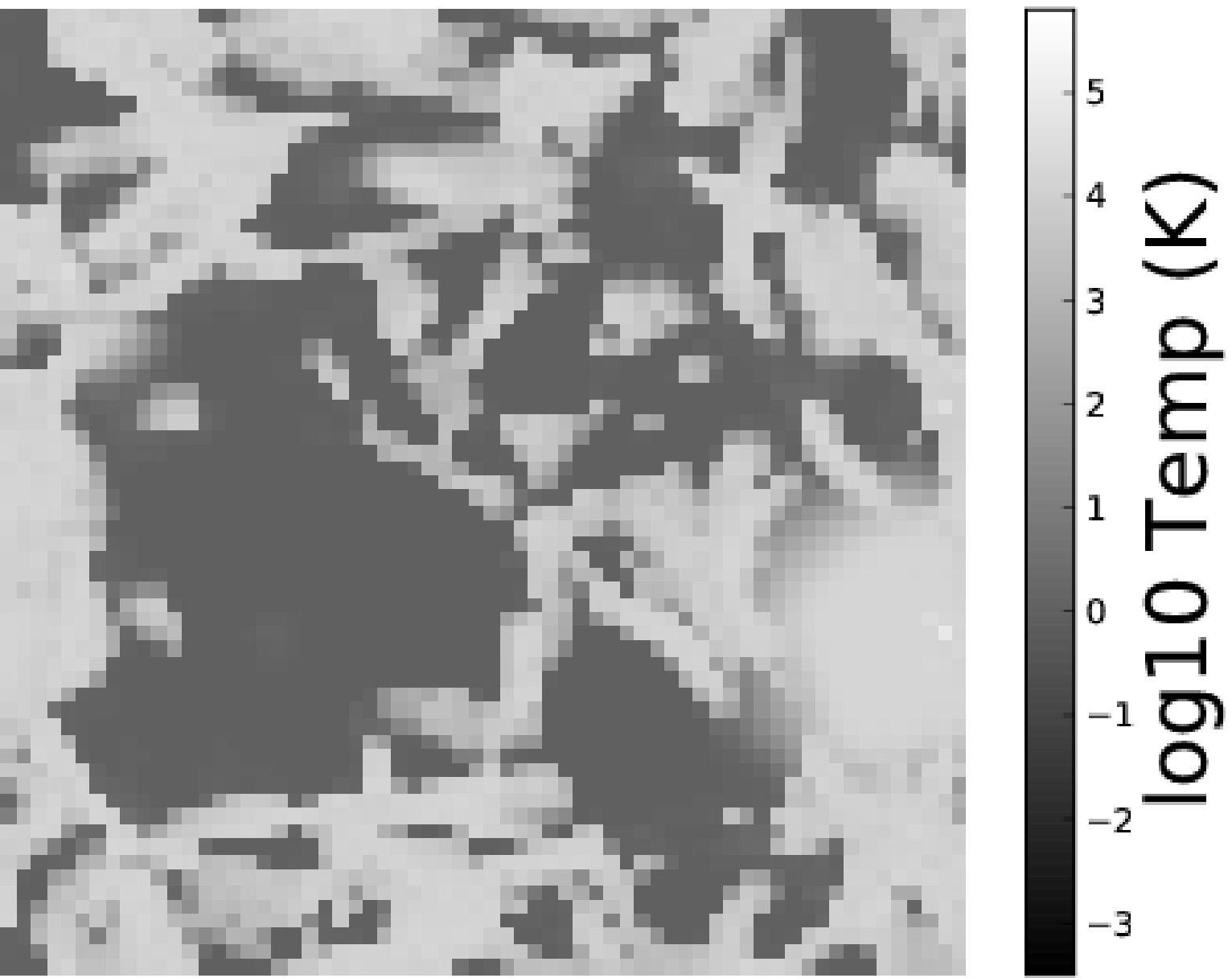} \\
    \includegraphics[width=0.3\textwidth]{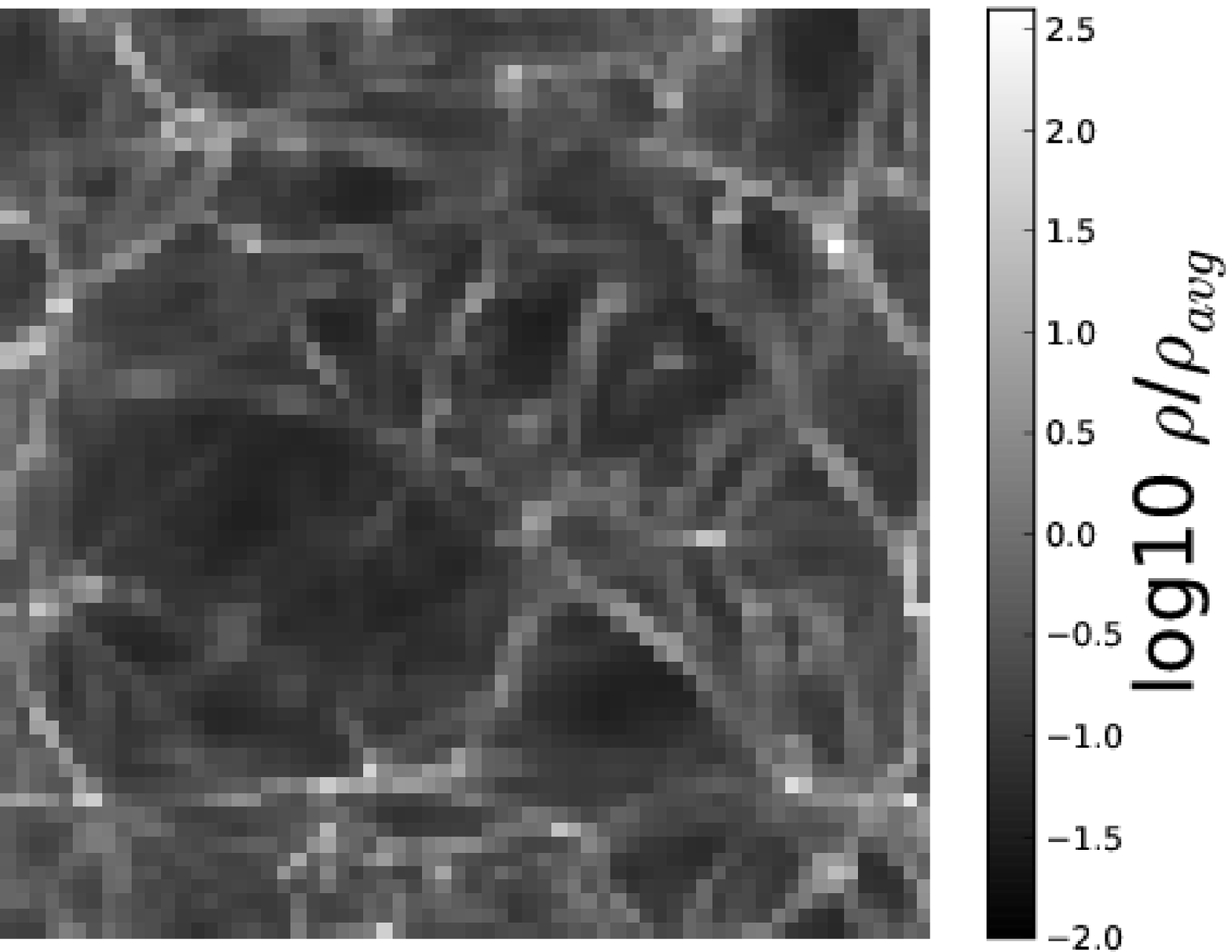} &
    \includegraphics[width=0.3\textwidth]{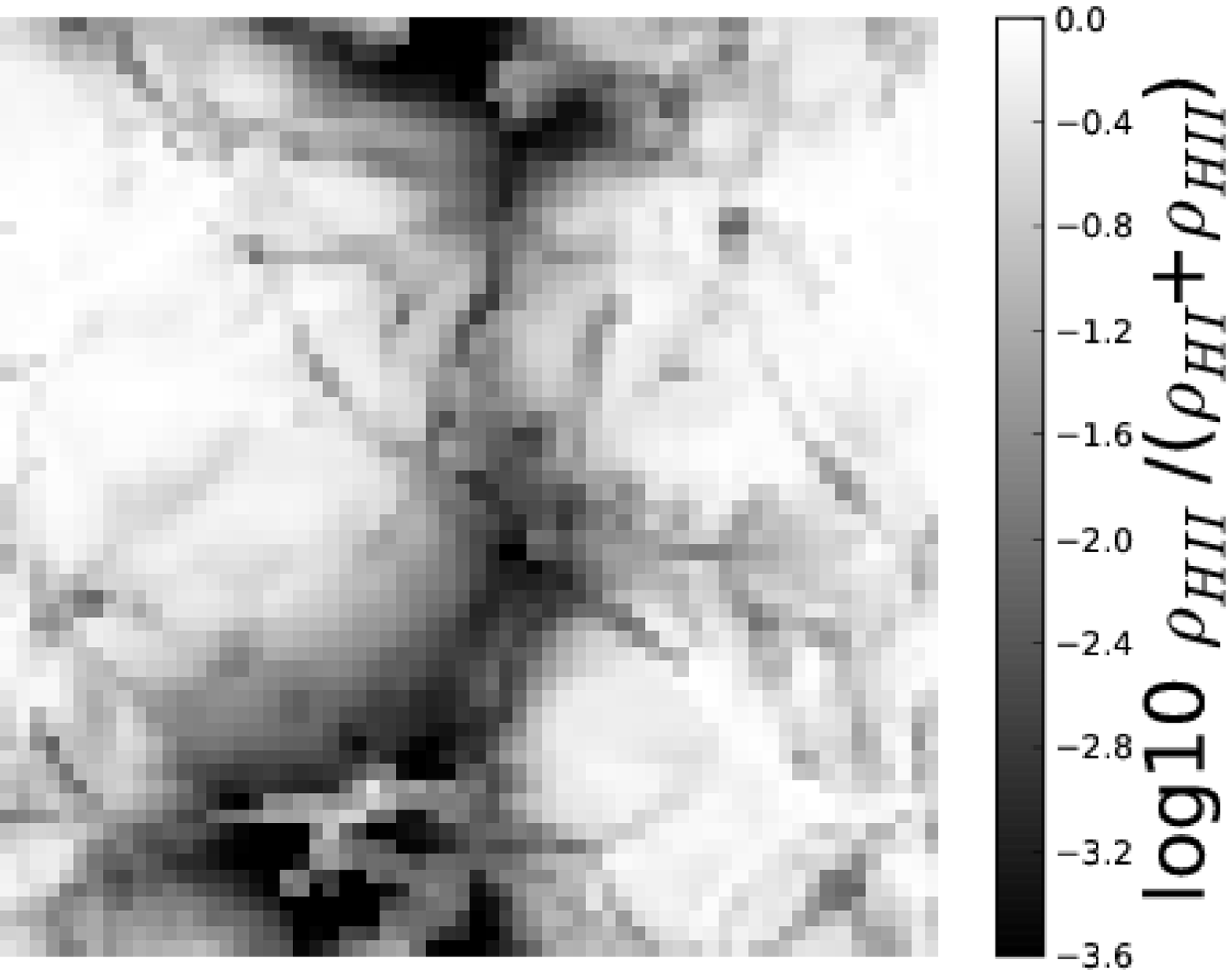} &
    \includegraphics[width=0.3\textwidth]{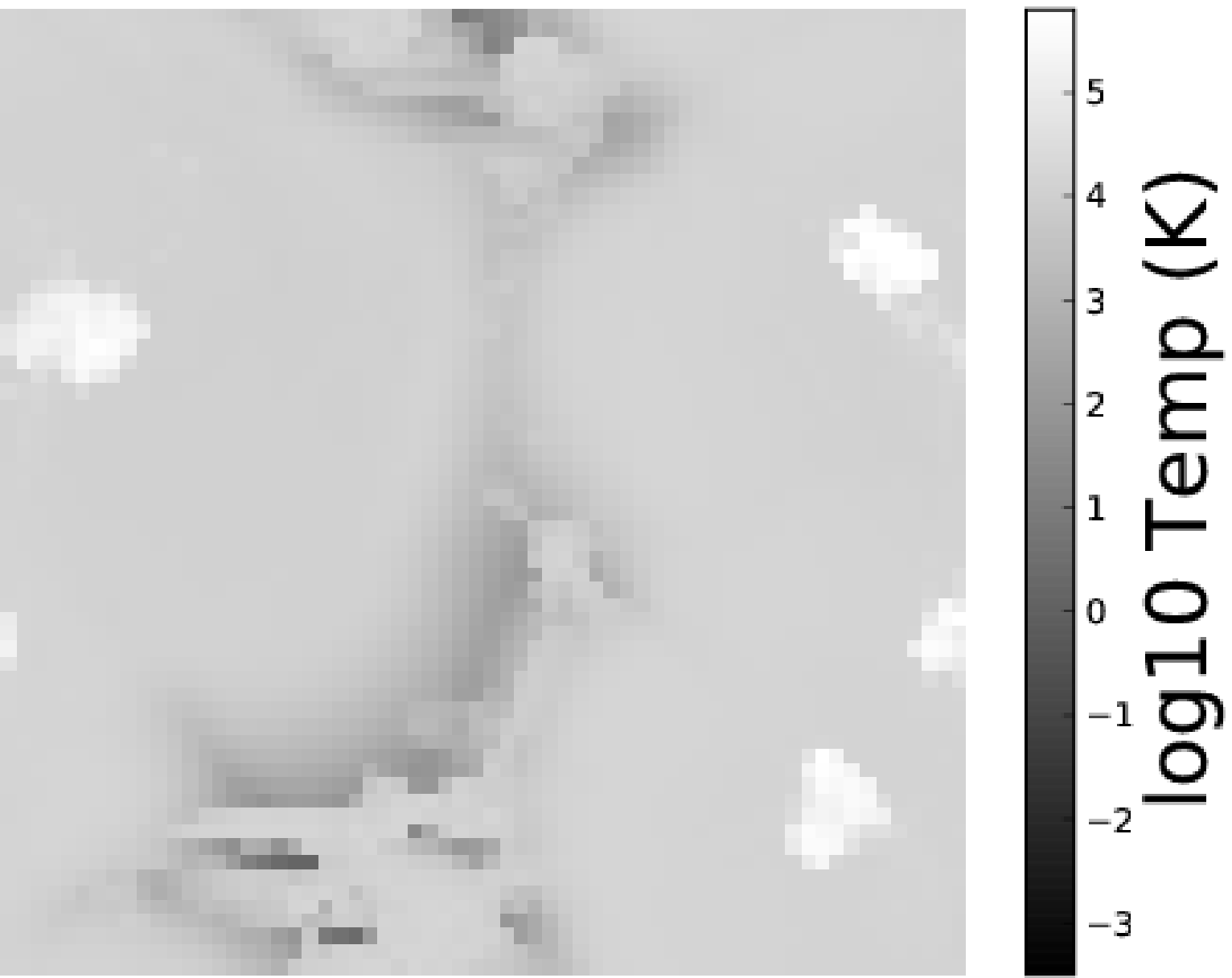} \\
    \includegraphics[width=0.3\textwidth]{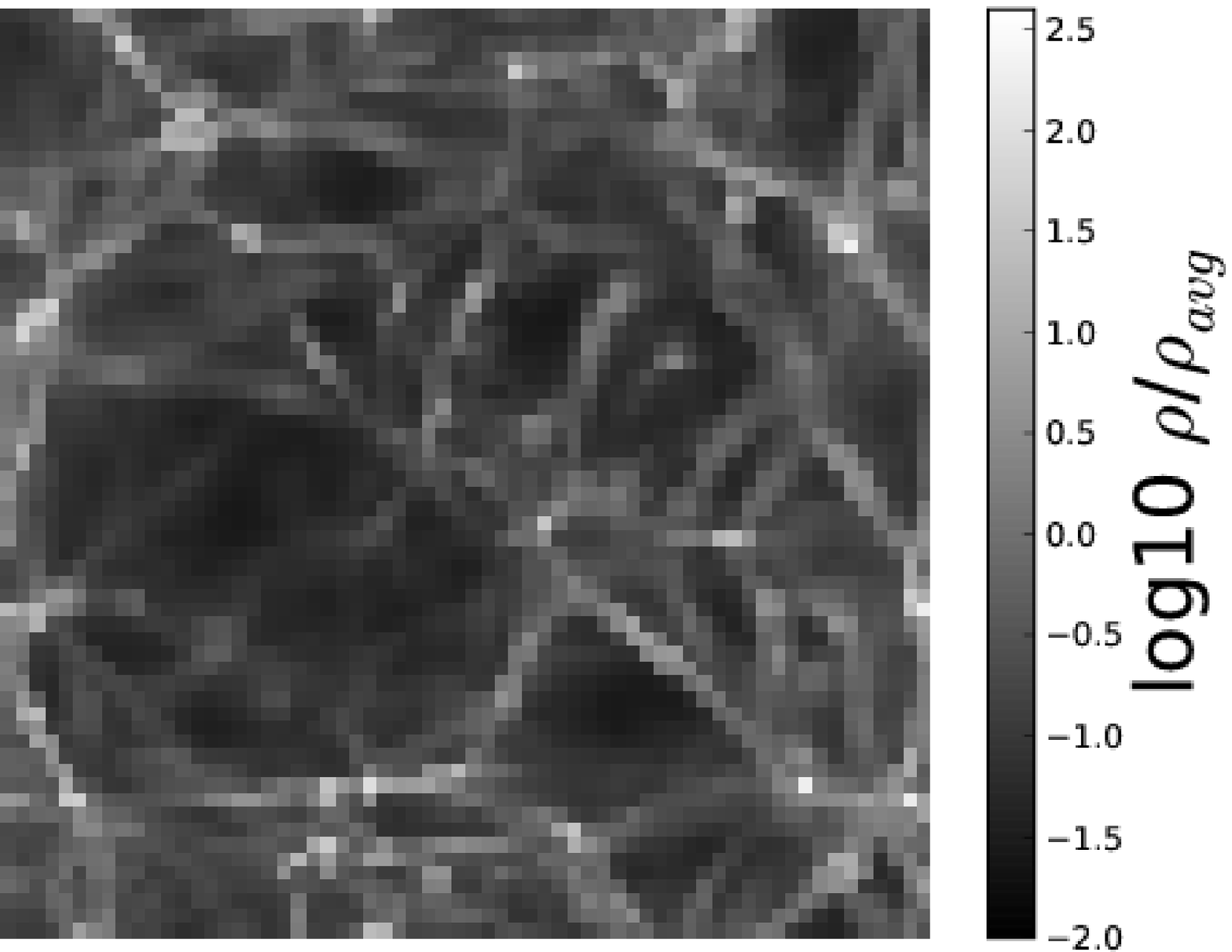} &
    \includegraphics[width=0.3\textwidth]{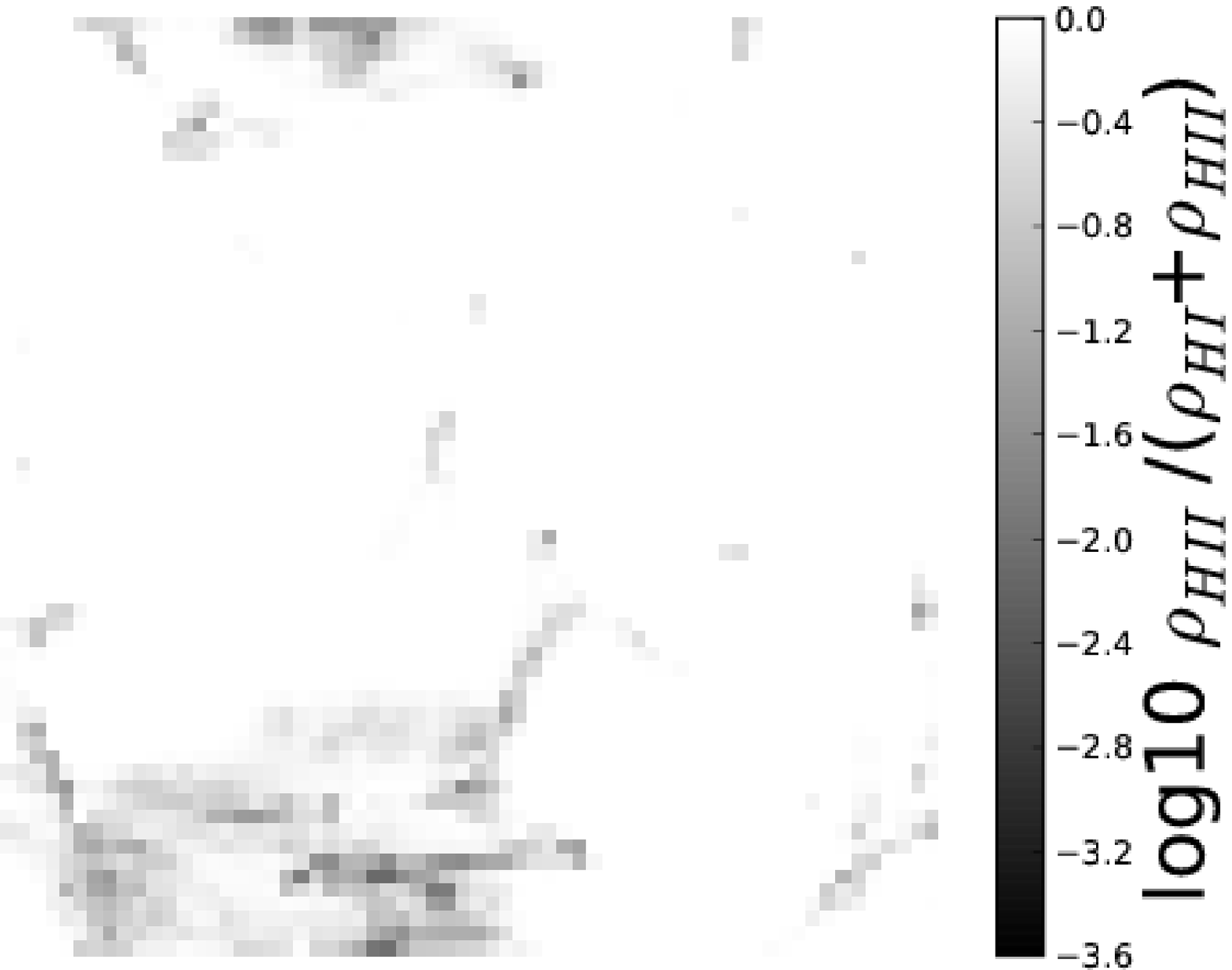} &
    \includegraphics[width=0.3\textwidth]{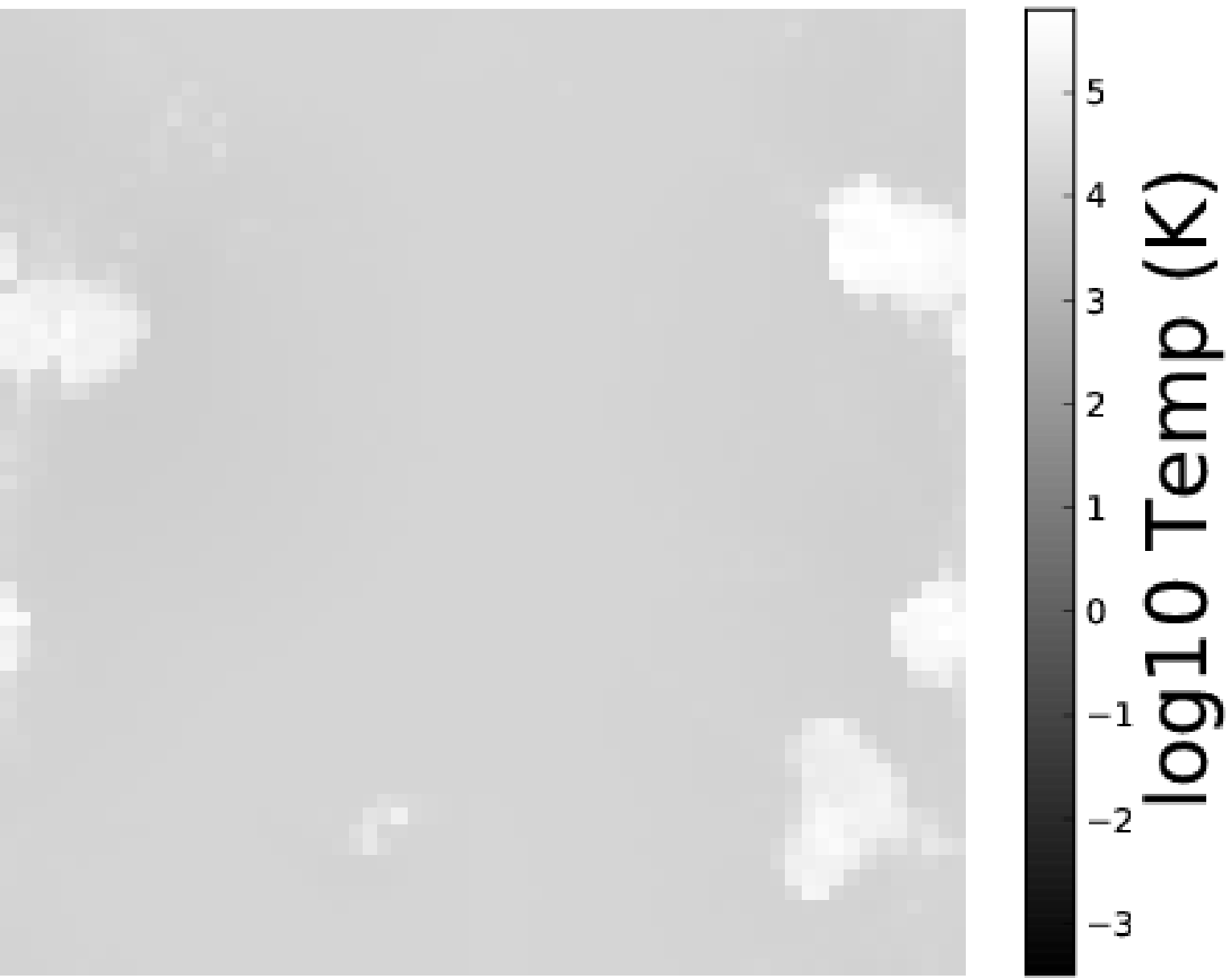} \\     
    \includegraphics[width=0.3\textwidth]{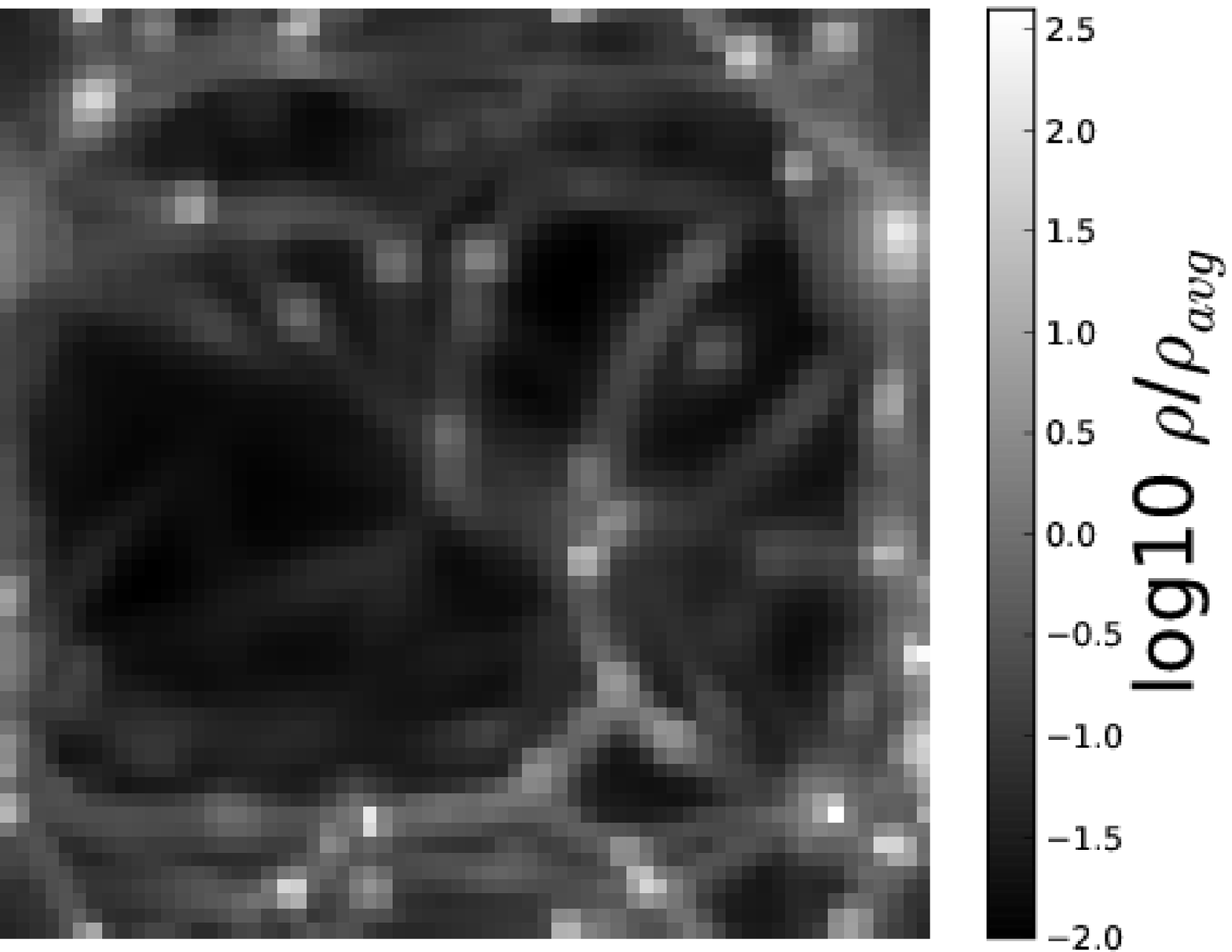} &
    \includegraphics[width=0.3\textwidth]{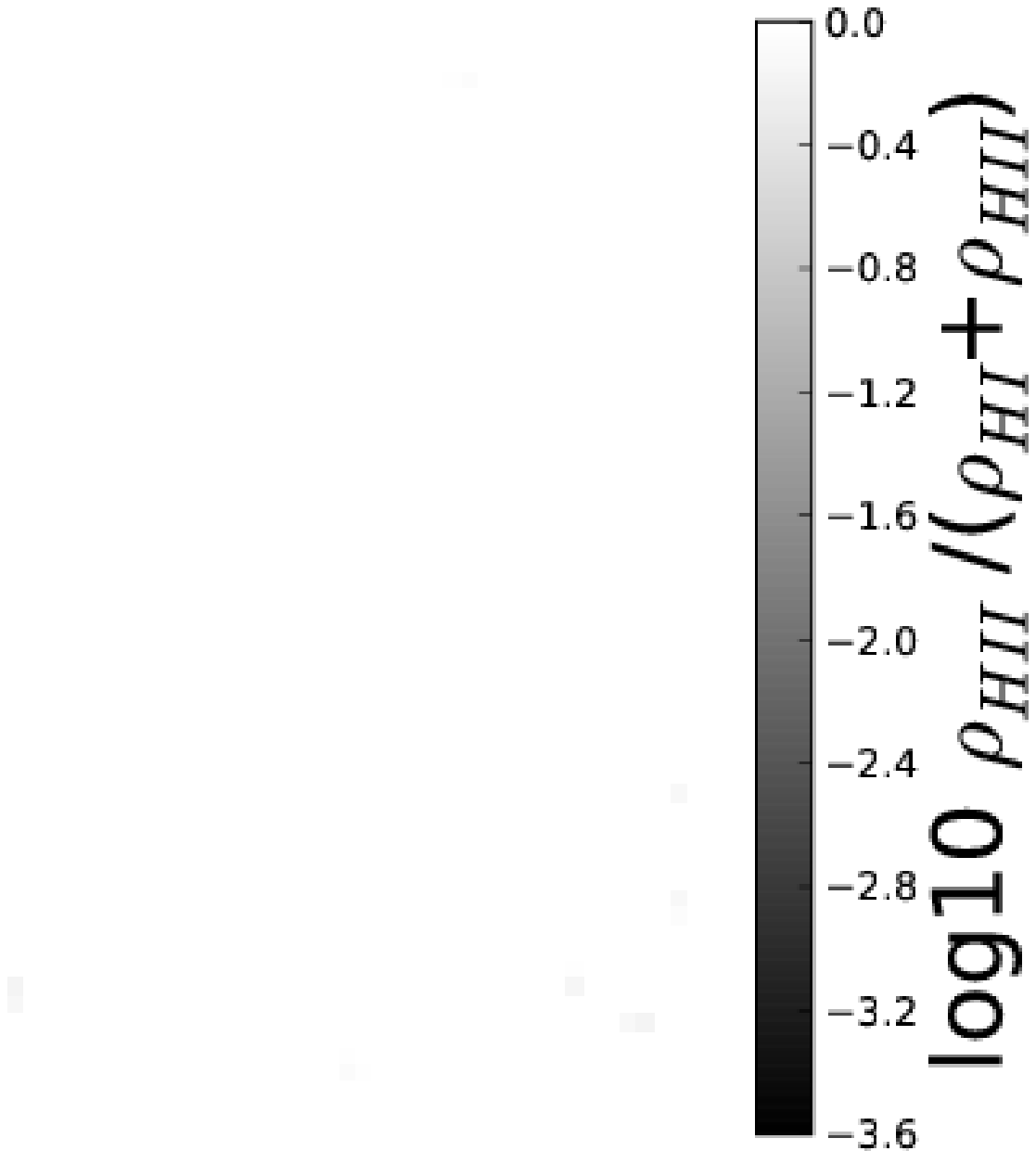} &
    \includegraphics[width=0.3\textwidth]{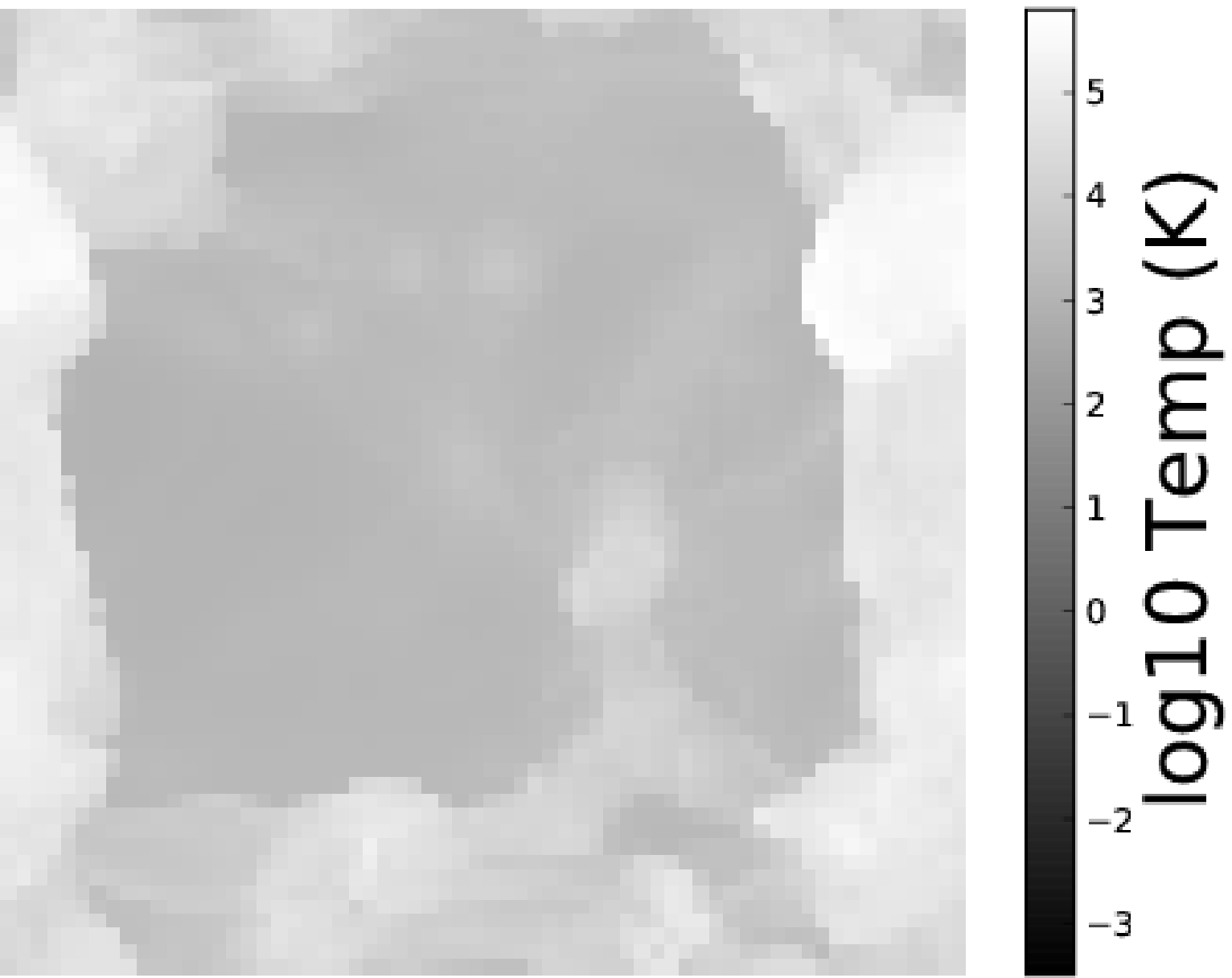} \\
    \end{tabular}
  \caption{Cosmological ionization snapshots: 2D data results from
    averaging through one direction. The rows correspond to times $z$
    = 4.35, 2.55, 1.99, 0; the columns show baryonic density,
    ionization fraction and temperature.} 
  \label{table:outputs}
\end{table}

By $z = 2.55$, multiple sources have formed and are contributing to
the ionizing radiation.  The ionization fronts have also clearly
propagated through significantly more of the domain.  Although the
ionization fronts are converging, there are still small pockets
where the IGM remains neutral.  By inspection, the majority of the IGM
is at around 10$^4$ K, consistent with expectation.  The bright peaks
in temperature mark a region with 4 hot stars in close proximity, but
the area of neutral IGM remained cooler. 

A short while later (on the cosmological scale) at $z = 1.99$,
there has been little change in the density structure, but the
ionization fronts have passed one other, overlapping the ionization
region.  At this point the universe is becoming transparent to
ionizing radiation.  Although in this simulation reionization has
finished much later than observed \cite{2006ARA&A..44..415F},
we note that the redshifts at which stars are made by this star maker
recipe are heavily dependent on the box size and spatial resolution of
the simulation.  A bigger box size with the same spatial resolution
will likely create stars at a much earlier redshift.  Meanwhile, most
of the computational volume has reached the same temperature, aside
from the local hot spots around the stars.

Finally at $z = 0$, the matter has nearly all coalesced due to the
gravitational potential of the high density peaks.  This results in
large voids of underdensity regions in the IGM, and at the same time
multiple bright spots are converging towards each other.  By this time
we see that the universe has been completely ionized (the data shows
very small specks where some HI exists, which may be attributed to
recombination).  Furthermore, the temperature plot shows that
most of the IGM is actually at a lower temperature than earlier at
reionization.  This is due to the adiabatic expansion of the universe,
causing regions far away from the sources of radiation to cool.  The
brighter temperature region is also expanding.  This is not due to the
photo-ionization of the IGM as before, but is instead due to
collisional heating from infall onto the massive dark matter halo, 
shock heating the regions around
it.  In a color plot, it can be seen that the shock front has a higher
temperature than the relaxation area behind the front. 

Our box is far too small, and the spatial resolution too low,
to describe the $z=0$ structure of the universe
accurately. Indeed, density fluctuations with wavelengths comparable to
the box size are going nonlinear at $z=0$, making our solution highly 
inaccurate. The sole purpose of continuing the calculation was to test
the long term stability of our implicit algorithm. It passed the test.

\section{Conclusions and outlook}

%<M.N.>

%[1 pg]
The combined code appears to be working as expected and is stable for long executions. 
Radiation from star formation fully ionizes the volume, however the redshift of reionization is delayed due to the low spatial
resolution which underestimates the star formtion rate. 
Higher resolution runs and larger box sizes are planned in the near future. As discussed in 
\cite{ReynoldsEtAl2009} our radiation solver is optimally scalable with respect to the number of radiation sources, the number of grid points, and the number of processors. 
Moreover, the timesteps for the radiation-ionization kinetics portion of the calculation is independent of resolution because of the implicit time differencing. This is not the case for explicit cosmological dynamics, which means that at some grid size the radiation portion of the calculation will cease to dominate the runtime cost. We have not yet determined where this crossover occurs, but are investigating the matter.

Several extensions of the method are under development. The first is multigroup FLD for a more accurate representation of the transport of hard UV and X-ray photons and helium ionization. A second is replacing the FLD ansatz with the variable tensor Eddington factor method used in \cite{2007arXiv0711.1904P}. This will improve the angular description of the radiation field and allow for shadowing effects. 

Finally, there is extending the radiation-ionization kinetics solver to adaptive mesh refinement. There are two components in this solver that depend on the spatial mesh.  The first of these is the solver for the Schur complement subsystem.  The part of the current solver for this component that currently depends on a uniform spatial mesh is the geometric multigrid solver that is used to precondition the conjugate gradient iteration.  In extending the approach outlined here to spatially adaptive meshes, this geometric multigrid solver may be replaced with a Fast Adaptive Composite (FAC) method that understands the overall composite mesh that is formed out of a nested hierarchy of uniform grids of different spatial resolution.

The second component that depends on a uniform spatial mesh is the rather straightforward operator-splitting approach coupling the explicit and implicit sub-solvers.  Due to the mesh-dependent CFL stability restriction, the explicit solvers employ time subcycling on the composite mesh, wherein more highly refined grids use smaller time steps than their larger counterparts, synchronizing with one another only at the time step of the coarsest grid.  The implicit solver, however, naturally couples all of these levels together at once.  Therefore in extending these solvers to AMR, we plan to examine the proper operator-splitting strategy for coupling these solvers together, attempting to balance a need for accuracy and consistency (use a full implicit solve every subcycled time step) with a need for efficiency (use a full implicit solve only at the coarsest grid time step).

%%%%%%%%%%%%%%%%%%%%%%%%%%%%%%%%%%%%%%%%%%%%%%%%
%% BACKMATTER
%%%%%%%%%%%%%%%%%%%%%%%%%%%%%%%%%%%%%%%%%%%%%%%%

\begin{theacknowledgments}
We would like to thanks our coauthors Pascal Paschos and John Hayes on \cite{ReynoldsEtAl2009} for their contributions to the formulation and testing
of method described here.
This work was supported in part by NSF grants AST-0708960 and AST-0808184. 
Simulations were performed on the Cray XT5 system {\em Kraken} at the National
Institute for Computational Sciences with NSF TeraGrid allocation TG-MCA98N020.
\end{theacknowledgments}



\bibliographystyle{aipproc}   % if natbib is available

\bibliography{sources}

\begin{thebibliography}{36}
\expandafter\ifx\csname natexlab\endcsname\relax\def\natexlab#1{#1}\fi
\providecommand{\enquote}[1]{``#1''}
\expandafter\ifx\csname url\endcsname\relax
  \def\url#1{\texttt{#1}}\fi
\expandafter\ifx\csname urlprefix\endcsname\relax\def\urlprefix{URL }\fi
\providecommand{\eprint}[2][]{\url{#2}}

\bibitem[{Reynolds} et~al.(2009)]{ReynoldsEtAl2009}
D.~R. {Reynolds}, J.~C. {Hayes}, P.~{Paschos}, and M.~L. {Norman}, \emph{ArXiv
  e-prints}  (2009), \eprint{0901.1110}.

\bibitem[{Fan} et~al.(2006)]{2006ARA&A..44..415F}
X.~{Fan}, C.~L. {Carilli}, and B.~{Keating}, \emph{\araa} \textbf{44}, 415--462
  (2006), \eprint{arXiv:astro-ph/0602375}.

\bibitem[{Ciardi} et~al.(2003)]{2003MNRAS.344L...7C}
B.~{Ciardi}, A.~{Ferrara}, and S.~D.~M. {White}, \emph{\mnras} \textbf{344},
  L7--L11 (2003), \eprint{arXiv:astro-ph/0302451}.

\bibitem[{Sokasian} et~al.(2004)]{2004MNRAS.350...47S}
A.~{Sokasian}, N.~{Yoshida}, T.~{Abel}, L.~{Hernquist}, and V.~{Springel},
  \emph{\mnras} \textbf{350}, 47--65 (2004), \eprint{arXiv:astro-ph/0307451}.

\bibitem[{Iliev} et~al.(2006)]{2006MNRAS.369.1625I}
I.~T. {Iliev}, G.~{Mellema}, U.-L. {Pen}, H.~{Merz}, P.~R. {Shapiro}, and M.~A.
  {Alvarez}, \emph{\mnras} \textbf{369}, 1625--1638 (2006),
  \eprint{arXiv:astro-ph/0512187}.

\bibitem[{Whalen} and {Norman}(2006)]{2006ApJS..162..281W}
D.~{Whalen}, and M.~L. {Norman}, \emph{\apjs} \textbf{162}, 281--303 (2006),
  \eprint{arXiv:astro-ph/0508214}.

\bibitem[{Iliev} et~al.(2009)]{2009arXiv0905.2920I}
I.~T. {Iliev}, D.~{Whalen}, G.~{Mellema}, K.~{Ahn}, S.~{Baek}, N.~Y. {Gnedin},
  A.~V. {Kravtsov}, M.~{Norman}, M.~{Raicevic}, D.~R. {Reynolds}, D.~{Sato},
  P.~R. {Shapiro}, B.~{Semelin}, J.~{Smidt}, H.~{Susa}, T.~{Theuns}, and
  M.~{Umemura}, \emph{ArXiv e-prints}  (2009), \eprint{0905.2920}.

\bibitem[{Bryan} et~al.(1995)]{BryanEtAl1995}
G.~L. {Bryan}, M.~L. {Norman}, J.~M. {Stone}, R.~{Cen}, and J.~P. {Ostriker},
  \emph{Comp. Phys. Comm.} \textbf{89}, 149--168 (1995).

\bibitem[{Hayes} and {Norman}(2003)]{HayesNorman2003}
J.~C. {Hayes}, and M.~L. {Norman}, \emph{Ap. J. Supp.} \textbf{147}, 197--220
  (2003).

\bibitem[{Paschos}(2005)]{Paschos2005}
P.~{Paschos}, \emph{On the Ionization and Chemical Evolution of the
  Intergalactic Medium}, Ph.D. thesis, University of Illinois at
  Urbana-Champaign (2005).

\bibitem[{Black}(1981)]{Black1981}
J.~{Black}, \emph{Mon. Not. R. Astr. Soc.} \textbf{197}, 553--563 (1981).

\bibitem[{Osterbrock}(1989)]{Osterbrock1989}
D.~E. {Osterbrock}, \emph{Astrophysics of Gaseous Nebulae and Active Galactic
  Nuclei}, University Science Books, Mill Valley, California, 1989.

\bibitem[{Cen}(1992)]{Cen1992}
R.~{Cen}, \emph{Ap. J. Supp.} \textbf{78}, 341 (1992).

\bibitem[{Abel} et~al.(1997)]{AbelEtAl1997}
T.~{Abel}, P.~{Anninos}, Y.~{Zhang}, and M.~L. {Norman}, \emph{New A.}
  \textbf{2}, 181--207 (1997).

\bibitem[{Hui} and {Gnedin}(1997)]{HuiGnedin1997}
L.~{Hui}, and N.~Y. {Gnedin}, \emph{MNRAS} \textbf{292}, 27--42 (1997).

\bibitem[{Razoumov} et~al.(2002)]{RazoumovEtAl2002}
A.~O. {Razoumov}, M.~L. {Norman}, T.~{Abel}, and D.~{Scott}, \emph{Ap. J.}
  \textbf{572}, 695--704 (2002).

\bibitem[{Cen} and {Ostriker}(1992)]{1992ApJ...399L.113C}
R.~{Cen}, and J.~P. {Ostriker}, \emph{\apjl} \textbf{399}, L113--L116 (1992).

\bibitem[{Hayes} et~al.(2006)]{HayesEtAl2006}
J.~C. {Hayes}, M.~L. {Norman}, R.~A. {Fiedler}, J.~O. {Bordner}, P.~S. {Li},
  S.~E. {Clark}, A.~{ud-Doula}, and M.-M.~M. {Low}, \emph{Ap. J. Supp.}
  \textbf{165}, 188--228 (2006).

\bibitem[{Hockney} and {Eastwood}(1988)]{HockneyEastwood1988}
R.~W. {Hockney}, and J.~W. {Eastwood}, \emph{Computer simulation using
  particles}, 1988.

\bibitem[{Norman} and {Bryan}(1999)]{NormanBryan1999}
M.~L. {Norman}, and G.~L. {Bryan}, \enquote{Cosmological Adaptive Mesh
  Refinement,} in \emph{Numerical Astrophysics}, edited by S.~M. {Miyama},
  K.~{Tomisaka}, and T.~{Hanawa}, 1999, vol. 240 of \emph{Astrophysics and
  Space Science Library}, p.~19.

\bibitem[{O'Shea} et~al.(2004)]{OSheaEtAl2004}
B.~W. {O'Shea}, G.~{Bryan}, J.~O. {Bordner}, M.~L. {Norman}, T.~{Abel},
  R.~{Harkness}, and A.~{Kritsuk}, \emph{Adaptive Mesh Refinement -- Theory and
  Applications}, Lecture Notes in Computational Science and Engineering,
  Springer, 2004, chap. Introducing Enzo, an {AMR} Cosmology Application.

\bibitem[{Colella} and {Woodward}(1984)]{ColellaWoodward1984}
P.~{Colella}, and P.~R. {Woodward}, \emph{J. Comp. Phys.} \textbf{54}, 174--201
  (1984).

\bibitem[{Norman} et~al.(2007)]{NormanEtAl2007}
M.~L. {Norman}, G.~L. {Bryan}, R.~{Harkness}, J.~{Bordner}, D.~R. {Reynolds},
  B.~{O'Shea}, and R.~{Wagner}, \emph{Petascale Computing: Algorithms and
  Applications}, CRC Press, 2007, chap. Simulating Cosmological Evolution with
  Enzo.

\bibitem[enz(????)]{enzo-site}
Enzo code project page, http://lca.ucsd.edu/portal/software/enzo (????).

\bibitem[{Reynolds}(2009)]{Reynolds2009}
D.~R. {Reynolds}, Improving the robustness and accuracy in simulating radiative
  chemical ionization (2009), manuscript in progress.

\bibitem[{Kelley}(1995)]{Kelley1995}
C.~T. {Kelley}, \emph{Iterative Methods for Linear and Nonlinear Equations},
  vol.~16 of \emph{Frontiers in Applied Mathematics}, SIAM, 1995.

\bibitem[{Knoll} and {Keyes}(2004)]{KnollKeyes2004}
D.~A. {Knoll}, and D.~E. {Keyes}, \emph{J. Comp. Phys.} \textbf{193}, 357--397
  (2004).

\bibitem[{Falgout} and {Yang}(2002)]{FalgoutYang2002}
R.~D. {Falgout}, and U.~M. {Yang}, \emph{Computational Science -- ICCS 2002
  Part III}, Springer-Verlag, 2002, vol. 2331 of \emph{Lecture Notes in
  Computer Science}, chap. hypre: a Library of High Performance
  Preconditioners, pp. 632--641.

\bibitem[hyp(????)]{hypre-site}
{HYPRE} code project page, http://www.llnl.gov/CASC/hypre/software.html (????).

\bibitem[{Baker} et~al.(2006)]{BakerFalgoutYang2006}
A.~H. {Baker}, R.~D. {Falgout}, and U.~M. {Yang}, \emph{Parallel Computing}
  \textbf{32}, 319--414 (2006).

\bibitem[{Shapiro} and {Giroux}(1987)]{ShapiroGiroux1987}
P.~R. {Shapiro}, and M.~L. {Giroux}, \emph{Ap. J.} \textbf{321}, L107--L112
  (1987).

\bibitem[Peebles(1993)]{Peebles1993}
P.~J.~E. Peebles, \emph{Principles of Physical Cosmology}, Princeton University
  Press, 1993.

\bibitem[{Trac} et~al.(2008)]{2008ApJ...689L..81T}
H.~{Trac}, R.~{Cen}, and A.~{Loeb}, \emph{\apjl} \textbf{689}, L81--L84 (2008),
  \eprint{0807.4530}.

\bibitem[{Eisenstein} and {Hu}(1999)]{1999ApJ...511....5E}
D.~J. {Eisenstein}, and W.~{Hu}, \emph{\apj} \textbf{511}, 5--15 (1999),
  \eprint{arXiv:astro-ph/9710252}.

\bibitem[Turk(2008)]{SciPyProceedings_46}
M.~Turk, \enquote{Analysis and Visualization of Multi-Scale Astrophysical
  Simulations Using Python and NumPy,} in \emph{Proceedings of the 7th Python
  in Science Conference}, edited by G.~Varoquaux, T.~Vaught, and J.~Millman,
  Pasadena, CA USA, 2008, pp. 46 -- 50.

\bibitem[{Paschos} et~al.(2007)]{2007arXiv0711.1904P}
P.~{Paschos}, M.~L. {Norman}, J.~O. {Bordner}, and R.~{Harkness}, \emph{ArXiv
  e-prints}  (2007), \eprint{0711.1904}.

\end{thebibliography}

\IfFileExists{\jobname.bbl}{}
 {\typeout{}
  \typeout{******************************************}
  \typeout{** Please run "bibtex \jobname" to optain}
  \typeout{** the bibliography and then re-run LaTeX}
  \typeout{** twice to fix the references!}
  \typeout{******************************************}
  \typeout{}
 }

\end{document}